%% file: draft_main.tex
\documentclass[12pt,english]{article}
\usepackage[utf8]{inputenc}
\usepackage[T1]{fontenc}
\usepackage{geometry}
\usepackage{color}
\usepackage{array}
\usepackage{float}
\usepackage{rotfloat}
\usepackage{booktabs}
\usepackage{multirow}
\usepackage{amsmath}
\usepackage{graphicx}
\usepackage{setspace}
\usepackage{epstopdf}
\usepackage{tikz}
\usepackage{pgfplots}
\usepgfplotslibrary{fillbetween}
\usetikzlibrary{patterns}
\usetikzlibrary{intersections}

\usepackage{tabularx}
\usepackage{subcaption}
\usepackage{threeparttable}
\usepackage[authoryear]{natbib}
\usepackage{comment}
\usepackage{amsfonts}
\usepackage{amssymb}
\usepackage{amsthm}

\makeatletter

\usepackage{lscape}

\usepackage{hyperref} 
\hypersetup{
    colorlinks=true, 
    linktoc=all,     
    linkcolor=blue,  
    citecolor=blue
}

\newtheorem{lem}{Lemma}
\newtheorem{prop}{Proposition}
\newtheorem{cor}{Corollary}

\newtheorem{defn}{Definition}

\newtheorem{asmp}{Assumption}

\usepackage[authoryear]{natbib}

\makeatletter

\author{\hspace{-1cm} Christopher Clayton\thanks{Yale School of Management and NBER; christopher.clayton@yale.edu.} \quad 
	Antonio Coppola\thanks{Stanford University Graduate School of Business and CEPR; acoppola@stanford.edu.\newline
    We thank Paul Fontanier, Stefano Giglio, Tarek Hassan, Arvind Krishnamurthy, Matteo Maggiori, Pablo Ottonello, and Jesse Schreger for helpful comments. Financial support from the Stanford GSB Business, Government, and Society (BGS) Initiative is gratefully acknowledged.} \hspace{0.0cm}
}

\usepackage{tikz}
\usetikzlibrary{positioning,arrows.meta,fit,shapes}
\usetikzlibrary{shapes.geometric, arrows.meta, positioning, calc, shadows, patterns, decorations.pathreplacing}
\usetikzlibrary{positioning, arrows.meta, shapes.multipart, backgrounds, calc}

\makeatother
\pgfplotsset{compat=1.18}

\begin{document}

\title{Financial Regulation and AI: A Faustian Bargain?}
\date{\medskip July 2025}
\maketitle
\onehalfspacing

\thispagestyle{empty}

\begin{abstract}
We examine whether and how granular, real-time predictive models should be integrated into central banks' macroprudential toolkit. First, we develop a tractable framework that formalizes the tradeoff regulators face when choosing between implementing models that forecast systemic risk accurately but have uncertain causal content and models with the opposite profile. We derive the regulator’s optimal policy in a setting in which private portfolios react endogenously to the regulator's model choice and policy rule. We show that even purely predictive models can generate welfare gains for a regulator, and that predictive precision and knowledge of causal impacts of policy interventions are complementary. Second, we introduce a deep learning architecture tailored to financial holdings data---a graph transformer---and we discuss why it is optimally suited to this problem. The model learns vector embedding representations for both assets and investors by explicitly modeling the relational structure of holdings, and it attains state-of-the-art predictive accuracy in out-of-sample forecasting tasks including trade prediction.
\end{abstract}

\vspace{0.3cm}
\vspace{2mm}


\newpage

\section{Introduction}

A central concern of macroprudential policy is identifying and mitigating the amplification of shocks through the financial system. Traditional macroprudential frameworks focus on ex ante regulation of well-established sources of financial fragility such as leverage accumulation or maturity mismatches. Yet over the past two decades, the data environment for regulators has changed dramatically. Supervisory filings and rapid data collection now give regulators the ability to observe granular, high-frequency information on investor portfolios. At the same time, predictive technologies such as deep learning have achieved significant gains in out-of-sample performance across domains. These advances raise the question: can real-time, high-dimensional predictive models be used productively in financial regulation and interventions? On the one hand, predictive models can help effectively detect signals of fragility that are not captured by canonical macroprudential metrics: where exactly fire sales may erupt, how crowded trades may unwind, or which asset classes are most exposed to redemption risk. On the other hand,  these models are often reduced-form and highly non-linear, with no guarantee that they might recover deep underlying structural forces that are invariant to the regulators' own use of the models.

This paper develops a theoretical and empirical framework to address this question. Our central object of analysis is the regulator's model choice and how it informs intervention. We ask whether regulators should deploy high-performing predictive models when they face uncertainty about the causal consequences of acting on the model's output. In this context, we show that the answer depends on the interaction between the predictive model's informational content as well as forecast accuracy, and the regulator's ability to target interventions and estimates of causal impact of those interventions. Empirically, we introduce and build new predictive architectures using deep learning models tailored to the relational structure of financial holdings data, and we show that these can achieve benchmark-setting predictive performance in the dimensions relevant to the macroprudential task, providing a blueprint for practical implementation by central banks and other regulators.

In our theory, we model a three-period economy with intermediaries and a government regulator. The model builds upon canonical fire sales environments. In the first period, financial intermediaries choose portfolios. In the second, intermediaries face constraints (e.g., pledgeability and collateral constraints) that force them to sell some assets prior to maturity. The key fire sale externalities arise because these assets are sold to second-best users whose productivity depends on the amount purchased. Finally, any assets held to maturity pay off in the third period.

The regulator can intervene in the second period to try to manage fire sales. We allow the regulator to employ a (potentially incomplete) set of liquidation wedges that affect intermediaries' choice of which assets to liquidate. Formally, liquidation wedges take the form of revenue-neutral taxes on selling an asset, although we later show that results extend to applying subsidies for retaining an asset. The key informational friction is that the regulator faces uncertainty over the mapping from intermediaries' asset positions and the policy intervention to realized liquidations and fire sale prices. The regulator has a prior over this mapping. Before undertaking the intervention, the regulator can choose to deploy a model that delivers a signal about the latent fundamentals (e.g., predicted sales volumes or liquidation discounts). The regulator can then design its intervention based on the model used and the signal produced. Crucially, models can differ in what aspects of the system they inform. A model may provide strong predictive content (about liquidations and prices) but little information about the causal effects of interventions, and vice versa.

We characterize the regulator’s optimal ex post policy rule as a function of its Bayesian posterior over key primitives. The regulator's optimal rule depends on the product between the predicted causal impact of the policy and the predicted social benefit of the policy. The social benefit of the policy, deriving from raising the fire sale price, is determined by the amount of each type of asset being sold. As a consequence, even a purely predictive model can improve welfare if its forecasts are aligned with dimensions of the system where the regulator has prior knowledge about the causal impact of the policy intervention. For example, if the regulator knows that its intervention will have a strong causal effect on fire sale prices in certain markets, then predictive models that are informative precisely about forced sales in those markets have particular potential to generate welfare gains. This suggests a complementarity between causal knowledge of policy interventions and predictive models that inform the social benefit of intervention.

We then characterize the expected welfare gains associated with the choice of a model by the regulator as well as optimal model choice. The welfare gains depend on the prior expectation of the size of the intervention, and also on the covariance matrix of the policy intervention (assessed from the prior perspective). For purely predictive models that are uninformative as to the causal structure, only this latter term depends on the choice of model. Hence choosing an optimal predictive model amounts to a choice of this covariance matrix. The intuition comes from the law of total variance. The regulator benefits from acquiring precise information on ex-post liquidations to better target its policies. However, not all predictive information is equally valuable. First, predictive precision is more valuable on margins where the regulator has prior knowledge that the policy intervention will have a strong causal impact, reinforcing the complementarity between causal knowledge and the value of predictive information. Second, the value of interventions (and hence of predictive precision) scales with the intervention's causal impact and with intermediaries' ex-post costs of portfolio adjustment. The regulator's optimal model choice thus tends to focus predictive precision on those margins with stronger predicted causal impacts and higher benefits of intervention.

To understand the implications of model deployment on intermediaries' decision making, we turn to the endogenous ex-ante portfolio response of private agents that anticipate the regulatory intervention. Intermediaries internalize the mapping from their portfolios to expected liquidation costs under the regulator’s anticipated policy, but take as given both what model the regulator will adopt and the intervention and fire sale prices. As a result, the regulator’s model choice affects portfolio selection. We show that the regulator's model choice and intervention can both discourage intermediaries from holding assets associated with fire sales ex post, but also potentially lead to moral hazard. Intuitively, the ex post intervention has two competing effects on the initial asset allocation. First, intermediaries directly perceive assets on which the regulator will impose liquidation wedges as more expensive to hold, and so shift portfolios away from these assets. This can create a virtuous effect of the ex-post intervention, potentially substituting for the need to regulate these assets ex ante. However, the countervailing effect is that by using the more informed intervention to raise liquidation prices ex post, the regulator reduces fire sale discounts and so encourages intermediaries to hold more of these assets. This latter effect can be particularly pronounced if the regulator is not able to regulate all holders of certain assets. All of these effects, both virtuous and moral hazard, are limited by the predictability and granularity of the model---that is, the precision with which the regulator can target fire sales in real time.

Further, we explicitly model a regulator's optimal ex-ante macroprudential intervention in the portfolio holdings of intermediaries. We allow the regulator to employ a set of wedges on asset holdings, modeled as revenue-neutral taxes. The regulator's optimal tax ex ante starts from the baseline regulation it would apply if it did not intervene ex post, and then makes three adjustments reflecting model choice and the ex post intervention. First, the regulator discourages holdings of assets that make it more costly to acquire precise predictive information ex post. Second, the regulator encourages relative holdings of assets that will cause it to increase the size of its ex-post intervention. Intuitively, the regulator recognizes for these assets that the larger ex-post regulation serves as a substitute for the ex-ante intervention, so the need to intervene ex ante is muted. Finally, the regulator reduces regulation of assets that are expected to be subject to high ex post taxes, since the ex post taxes help discourage ex ante purchases of those assets. Conversely, the regulator increases regulation of assets for which moral hazard is induced by ex post bolstering of asset prices.

Although we illustrate our key insights for policy interventions in the form of liquidation wedges, in practice ex post interventions often resemble ``bailout'' measures with a subsidy component. We accommodate this policy in an extension in which we assume the regulator ex post imposes an asset retention wedge, formally modeled as a revenue-neutral subsidy for holding an asset to maturity. These subsidies imply a de facto tax on liquidating an asset, and accordingly the optimal intervention and model design remain the same. However, these subsidies introduce an additional moral hazard channel, since assets that are subsidized ex post become more attractive to purchase ex ante. Because this moral hazard effect depends on the expected subsidy size, models that increase predictive precision but do not change the expected tax rate are not subject to it (akin to \citealt{laffont1986}). Since in our setup purely predictive models increase precision without changing the expected intervention size, the moral hazard they generate interestingly does not differ relative to the case with liquidation taxes instead of asset retention subsidies.

To empirically assess whether predictive models can in fact deliver useful signals for these macroprudential purposes, we develop a deep learning architecture tailored to the structure of financial holdings data. Much of the modern deep learning toolkit---including models that have been applied in asset pricing contexts---is tailored toward grid and sequence inputs such as images and text. Yet in domains where the data has important graph structure, architectures tailored specifically to learn from graph structures have achieved breakthrough performance, including for instance when applied to problems of protein folding prediction and drug discovery (\citealt{jumper2021highly}). In our setting, the data naturally form a graph: investors are connected to assets through their positions. We therefore design our empirical deep learning models to explicitly capture and make use of this crucial dimension of the data.

To exploit the relational structure of holdings data, we implement and train a \textit{graph transformer} model, a form of graph neural network (GNN) augmented with an attention mechanism, to learn latent representations of both investors and assets (i.e., investor and asset embeddings into latent vector spaces). The architecture features two properties that are essential for this setting. First, it is permutation-invariant: predictions do not depend on the arbitrary ordering of investors or assets. This is a key distinction with sequence-based architectures (such as sequence transformers which underpin modern large language models) commonly used for text, in which the order of words in a document is essential to their contextual meaning. Second, the graph transformer model is inductive: all model parameters are fully shared across nodes, allowing the model to generalize to new investors or assets without retraining, while at the same time enforcing a strong form of regularization that leads to good generalization out of the model's training sample. The model is optimally sample-efficient in the sense of requiring minimal degrees of freedom to learn functions of the holdings data that are themselves invariant to arbitrary relabelings of investors and assets---a restriction that reflects the economics of the problem. Intuitively, this occurs because graph-based architectures do not need to use parameters to relearn permutation invariance from the data, but rather they enforce it explicitly through their structure.

We train the model jointly on two tasks: a masked autoencoder task, in which the model learns to reconstruct partially masked holdings, and a supervised prediction task, in which the model forecasts the cross-sectional pattern of future trades. Both tasks are designed to inform the model about latent economic relationships governing portfolio choice and rebalancing. The training generates embeddings for both assets and investors which are general-purpose and can be used for multiple objectives. We do not train the model explicitly to predict fire sales in a supervised fashion, but rather turn to models which can learn general-purpose embeddings, precisely to avoid in-sample over-fitting and thus avoid a sharp degradation in out-of-sample performance. The training uses quarterly holdings data from Factset, which cover a broad range of institutional investors and assets.

We find that the model performs well on both tasks. It accurately reconstructs positions data with correlations exceeding 90\% between predictions and targets, and it achieves strong out-of-sample predictive accuracy of nearly 30\% in forecasting trade patterns.  Notably, the model's performance is indeed stable out of sample, due to its inductive structure and shared-parameter design. The high holdings reconstruction fidelity on the autoencoder task should naturally be interpreted in light of the model's parameter-to-data ratio: with roughly 3.6 million parameters---representing less than 1\% of the possible asset-investor pairs in the holdings data---the model extracts economically relevant patterns from high-dimensional data. Further, we show that the model's performance in forecasting trading behavior (our second task) remains high even when restricting the sample to the set of active investment managers only and to stress periods, both of which are especially relevant to the macroprudential application we are studying. For example, we show that the model is able to forecast the pattern of asset trades during the market crash of 2020 induced by the Covid pandemic with good accuracy, having been trained only on pre-2020 data.

The predictive success of our deep learning architecture provides empirical support for the potential of model-informed ex post regulation. Moreover, we view our model as a blueprint for real-time regulatory approaches: a framework to transform high-frequency, granular data into relatively low-dimensional representations suitable for intervention design. Although the model is not structurally interpretable in the traditional sense, our theory shows that it can nonetheless play a useful role when paired with prior knowledge of structural effects.

\paragraph{Related Literature.} 
At their heart, our questions are situated within a longstanding intellectual debate concerning whether models that are primarily predictive in nature should be used to inform macroeconomic policy and financial regulation. \cite{koopmans1947measurement}, representative of the viewpoints then prevalent at the Cowles Commission, introduced his landmark ``measurement without theory'' critique of the earlier empirical studies of business cycles by \cite{burns1946measuring}. These early debates set the stage for the methodological ideas of \cite{haavelmo1944probability} and \cite{friedman1953methodology}, and for \textcolor{blue}{\citeauthor{lucas1976econometric}'s }(\citeyear{lucas1976econometric}) eventual critique of policy evaluation carried out without individual-optimization microfoundations. While the qualitative outline of this debate has remained the same, its quantitative content has evolved, as the performance and generalization ability of predictive models has increased sharply. In this context, we show that sufficiently capable predictive models can have a role in the macro-regulatory toolbox, as a complement (rather than a substitute) to traditional structural approaches.

Our analysis relates to four broad areas of the literature. First, we connect to the literature on fire sales, and macroprudential regulation and ex-post interventions. Theoretical contributions include \cite{bernanke1986agency}, \cite{williamson1988corporate}, \cite{kiyotaki1997credit}, \cite{Caballero2001}, \cite{Lorenzoni2008}, \cite{Bianchi2011,bianchi2016}, \cite{stein2012}, \cite{farhi2016theory}, \cite{charikehoe2016}, \cite{bianchimendoza2016}, \cite{davilakorinek2018}, and \cite{claytonschaab2022,claytonschaab2025}. On the empirical side, prior work has focused on particular variables to explain and predict the occurrence and magnitude of fire sales, such as leverage (\citealt{fisher1933debt}; \citealt{kindleberger1978manias}; \citealt{brunnermeier2013bubbles}; \citealt{schularick2012credit}; \citealt{adrian2014procyclical}; \citealt{krishnamurthy2025credit}) and investor composition (\citealt{brainard1968pitfalls}; \citealt{coval2007asset}; \citealt*{haddad2021selling}; \citealt{coppola2025safe}; \citealt*{fang2025holds}): the empirical contribution of this paper asks whether there is additional value from a more agnostic approach that does not impose an ex-ante focus on particular dimensions of the data.

Second, we connect to the literature on applications of machine learning and deep learning to finance. \cite*{gu2020empirical, gu2021autoencoder}, \cite*{kozak2020shrinking}, \cite{nagel2021machine}, and \cite*{bryzgalova2023asset} apply machine learning techniques to the canonical empirical asset pricing problem of valuation in the cross-section of assets. Similarly, \cite*{chen2024deep} introduce a deep learning architecture for modeling stock returns. \cite*{giglio2022factor} review the intersection of empirical asset pricing and machine learning. \cite*{gabaix2024asset} use holdings data together with a variety of deep learning sequence models (e.g., Word2Vec and BERT) and recommender systems to estimate asset embeddings, and \cite*{gabaix2025upgrading} apply these to explain variation and volatility in credit spreads. \cite*{dolphin2022stock}  and \cite{sarkar2025economic} also construct asset embeddings, respectively from return data and textual sources. Methodologically, our paper employs graph neural network architectures (\citealt*{scarselli2008graph}; \citealt*{hamilton2017inductive}; \citealt*{xu2018powerful}; \citealt*{wu2020comprehensive}) to learn representations from the relational structure of asset holdings data.\footnote{See also \cite*{elliott2014financial} and \cite*{acemoglu2015systemic} for theory highlighting the importance of the network structure of positions for financial stability.}

Third, we relate to literature studying the deployment of machine learning and large-scale data for economic analysis and policy problems more broadly, including \cite{einav2014economics, einav2014data}, \cite*{ kleinberg2015prediction}, \cite{athey2017beyond, athey2018impact}, \cite{mullainathan2017machine}, \cite*{kleinberg2018human}, \cite{gillis2019big}, \cite{farboodi2020long}, and \cite{athey2021policy}. The question of how non-policy invariant relationships should be exploited by regulators also has earlier conceptual parallels, for instance by \cite{barro1983positive} in the context of the Phillips curve. Lastly, the model and information design aspect of our theoretical approach connects to work on stress testing (\citealt*{shapiro2015information, faria2017runs, goldstein2018stress, leitner2023model, orlov2023design,  parlatore2025designing}).

\section{A Framework for Regulatory Model Choice}
There are $N$ assets. There are $I$ intermediary types (each of equal measure), with a representative intermediary of each type $i$. There is also a representative arbitrageur. The model has a Beginning-Middle-End structure. In the Beginning, initial asset positions are undertaken. In the Middle, intermediaries may be forced to sell assets prior to maturity to arbitrageurs, who are second best users. In the End, payoffs are distributed and consumption occurs.

\paragraph{Intermediaries.} Intermediaries are risk neutral. In the Beginning, intermediary $i$ invests in a vector $q_i=(q_{i1},\ldots,q_{iN})^T$ of assets. If $q_{in}<0$, then intermediary $i$ has undertaken a negative investment (shorting) asset $n$.\footnote{We could endow intermediaries with a stock of assets, but for expositional convenience we set that stock to $0$.} The cost to intermediary $i$ of producing the asset vector $q_i$ is $C_i(q_i) = p_i^T q_i - \frac{1}{2} q_i^T H_i^q q_i$, where $p_{in}$ is the per-unit cost with $p_i = (p_{i1},\ldots,p_{iN})^T$. The cost component $q_i^T H_i^q q_i$ is a quadratic adjustment/holding cost, where $H_i^q$ is an $N\times N$ matrix. If held to maturity, intermediary $i$'s holdings of asset $n$ will produce a per-unit return $R_{in}$ in the End, with $R_i = (R_{i1},\ldots,R_{iN})^T$.

In the Middle, intermediary $i$ can sell assets prior to maturity. We denote $\ell_{in}\geq 0$ to be sales by intermediary $i$ at endogenous price $\gamma_n$, with $\ell_{i} = (\ell_{i1},\ldots,\ell_{iN})^T$ and $\gamma = (\gamma_1,\ldots,\gamma_N)^T$. Intermediary $i$ faces an adjustment cost $\frac{1}{2}\ell_{i}^{T}H_{i}^
\ell\ell_{i}$ on asset sales, where $H_i^\ell$ is $N\times N$. Assets will be sold at a discount on their fundamental value ($\gamma < R_i$, see below), and intermediary $i$ faces a set of ``rollover constraints'' that force asset sales. This set of $M$ constraints is given by
\begin{equation}\label{eqn:rollover}
A_{i}^q q_{i} + \rho_{i} \leq A_i^\ell\ell_i,
\end{equation}
where $A_i^q,A_i^\ell$ are $M\times N$ and $\rho_i$ is $M\times 1$. For example, equation \ref{eqn:rollover} can capture constraints on asset positions (e.g., a limit on debt) or a requirement to raise funds based on asset holdings (e.g., a cost of maintaining the project).\footnote{We simplify analysis by not having equation \ref{eqn:rollover} depend on the liquidation price $\gamma$, as for example in price-dependent collateral constraints. This enables us to maintain a linear-quadratic structure throughout the paper. It is straightforward to extend analysis to include prices in constraints, but the characterization of the ex-ante optimal portfolio would no longer admit a closed-form solution.}

In the End, intermediary $i$ realizes payoff on assets held to maturity and consumes. Intermediary $i$'s total payoff (consumption) in the End, inclusive of adjustment costs, is
\begin{equation}\label{eqn:intermediary_payoff_end}
    U_{i}=q_{i}^{T}(R_{i}-p_{i})-\ell_{i}^{T}(R_{i}-\gamma)-\frac{1}{2}q_{i}^{T}H_{i}^{q}q_{i}-\frac{1}{2}\ell_{i}^{T}H_{i}^{\ell}\ell_{i}.
\end{equation}

\paragraph{Arbitrageurs.} A representative arbitrageur is a second-best user of intermediary assets. If the arbitrageur purchases a vector $L=(L_{1},\ldots,L_N)^T$ of intermediary assets in the Middle, the arbitrageur can use them in a production technology to produce $\mathcal F(L) = L^T\overline\gamma - \frac{1}{2}L^T\Gamma L$ units of the consumption good in the End, where $\overline\gamma$ is $N\times 1$ and $\Gamma$ is $N\times N$. The representative arbitrageur's payoff is
\begin{equation}\label{eqn:arbitrageur_payoff}
U_A = L^{T}(\overline\gamma-\gamma)-\frac{1}{2}L^{T}\Gamma L.
\end{equation}

\paragraph{Information Structure and Timing.} Although all model parameters are determined in the Beginning, all agents in the Beginning are uncertain about the true values of the parameters $\Phi = \{A_i^q,A_i^\ell,H_i^\ell,\rho_i,R_i,\overline\gamma\}$. That is, agents are uncertain as to the true parameters underlying the rollover constraint (equation \ref{eqn:rollover}), the asset return $R_i$, the arbitrageurs' baseline productivity $\overline\gamma$, and the liquidation adjustment cost $H_i^\ell$. All agents have a common prior $\Phi\sim\mu_0$ in the Beginning. In the Middle before asset sales and purchases are chosen, the true model parameters become common knowledge of private agents. After becoming common knowledge, intermediaries choose asset liquidations and arbitrageurs choose asset purchases.\footnote{Assuming that private agents and the regulator (see below) have a common prior in the Beginning simplifies analysis because it prevents the regulator from learning information about these parameters from inference about private sector beliefs based on the asset allocation. It also eliminates a regulatory incentive based on different beliefs the regulator and agents (e.g., \citealt{fontanier2025}.)

Assuming $\Gamma$ is known to all agents ex ante simplifies analysis by eliminating the ability of a regulator to learn about the price impact of liquidations from observing a signal of $\gamma$ or $\ell$. This will allow us to fully separate predictive and causal channels in our framework. Absent this assumption, predictive models that inform a regulator about $\gamma$ and $\ell$ might have even more value ex post because they would also allow the regulator to learn about these structural parameters that determine the causal impact of a policy intervention (even if very little information is learned).}

\paragraph{Market Clearing.} Markets must clear for liquidations in the Middle, that is
\begin{equation}\label{eqn:marketclearing}
L = \sum_{i} \ell_i.
\end{equation}

\paragraph{Competitive Equilibrium.} We define a competitive equilibrium as follows.

\begin{defn}
A competitive equilibrium of the model, given true model parameters $\Phi$, is a vector of prices $\gamma$ and a set of allocations $\{q_i,\ell_i,L\}$ such that:
\begin{enumerate}
    \item In the Middle, taking as given asset allocations $\{q_i\}$ and model parameters $\Phi$:
    \begin{enumerate}
    \item Intermediary $i$ chooses $\ell_i$ to maximize utility (equation \ref{eqn:intermediary_payoff_end}) subject to the rollover constraint (\ref{eqn:rollover}), taking prices $\gamma$ as given.
    \item Arbitrageurs choose $L$ maximize utility (equation \ref{eqn:arbitrageur_payoff}), taking prices $\gamma$ as given.
    \item The liquidation markets clear (equation \ref{eqn:marketclearing}).
    \end{enumerate}
    \item In the Beginning:
\begin{enumerate}
    \item Intermediary $i$ chooses $q_i$ to maximize expected utility $\mathbb E_0[U_i]$, where $\mathbb E_0$ denotes the expectation given the prior $\mu_0$ over model parameters $\Phi$.
\end{enumerate}
\end{enumerate}
\end{defn}

\subsection{Regulator's Model Design and Intervention in the Middle}
In the Middle, a regulator is able to intervene in order to try to manage the fire sale price impact of liquidations. Formally, the regulator can impose a vector $\tau_i=(\tau_{i1},\ldots,\tau_{iN})^T$ of revenue-neutral liquidation wedges on each intermediary $i$, which alter the intermediary's perceived price for selling the asset.\footnote{In Section \ref{sec:subsidy}, we instead assume the regulator must use asset holding subsidies rather than liquidation taxes. The results in the Middle are identical, but the asset holding subsidies introduce additional moral hazard in the Beginning.} $\tau_{in}$ represents a tax on selling asset $n$, with $\tau_{in}<0$ being a subsidy for sale. As a result, intermediary $i$'s payoff in the End is modified to be
\begin{equation}\label{eqn:wedges}
U_i - (\ell_i^T - \ell_i^{\ast T})\tau_i
\end{equation}
where $\ell_i^{\ast T}\tau_i$ is equilibrium revenue remissions based on the equilibrium asset liquidations $\ell_i^{\ast}$ of intermediary $i$. Intermediaries take revenue remissions as given. We allow for the possibility that the regulator has potentially incomplete instruments, represented by a regulatory cost $\delta(\tau) = \frac{1}{2}\tau^T \Delta \tau$, where $\Delta$ is $NI\times NI$.\footnote{One could instead model incompleteness as a restriction $\Delta(\tau)\leq 0$ in which case the regulator's Lagrangian will be $\mathbb E[\sum_i U_i|\Gamma,M]-\lambda\Delta(\tau)$ where $\lambda$ is the Lagrange multiplier. This is akin to the reduced-form cost $\delta(\tau)=\lambda\Delta(\tau)$ except that $\lambda$ is endogenous.}

Unlike private agents, the regulator does not learn the true model parameters $\Phi$. Instead, the regulator has to use a \textit{model} to make an inference about the true parameters. The regulator's \textit{model} $M\in\mathcal M$ formally is a process of drawing a signal $s$ of the parameters. The signal updates the regulator's posterior distribution to $\Phi\sim \mu_p|s,M$. The regulator must choose the wedges $\tau$ before private agents move, and so the regulator cannot obtain any new information from observing the market before setting regulation (apart from the signal drawn).

To simplify exposition, we introduce an assumption of matrix symmetry that we maintain throughout the paper.
\begin{asmp}\label{asmp:symmetry}
    The matrices $H_i^q,H_i^\ell,\Gamma,\Delta$ are symmetric.
\end{asmp}

\paragraph{Impact of Regulatory Intervention.} To solve the regulator's optimum, we begin by characterizing the impact of the regulator's wedges $\tau = (\tau_1^T,\ldots,\tau_I^T)^T$ on the equilibrium in the Middle. Our model is substantially simplified by the information structure: because private agents directly observe $\Phi$ in the Middle, for a given vector of asset allocations $q=(q_1^T,\ldots,q_I^T)^T$ the equilibrium in the Middle does not depend on the regulator's choice of model $M$ or the realized signal $s$ except through the choice of wedges $\tau$.

The linear-quadratic structure of preferences and the rollover constraint leads to a characterization of the equilibrium as a simple linear system of equations. The following Lemma characterizes this equilibrium.

\begin{lem}\label{lem:equil}
Given asset allocations $q$, true parameters $\Phi$, and regulatory wedges $\tau$, the equilibrium in the Middle is given by
\begin{equation}\label{eqn:equil_ell}
\ell_{i}=\overline{\ell}_{i}+\Lambda_{i}^{q}q_{i}-\Lambda_{i}^{\tau}\tau_{i}-\sum_{j=1}^N\bigg[\Lambda_{j}^{q,e}q_{j}^{\ast}-\Lambda_{j}^{\tau,e}\tau_{j}^{\ast}\bigg]
\end{equation}
\begin{equation}\label{eqn:equil_gamma}
\gamma=\overline{\gamma}-\Gamma\sum_{i}\ell_{i}
\end{equation}
where $\overline\ell_i$ ($N\times 1$) and $\Lambda_i^q,\Lambda_i^\tau,\Lambda_i^{q,e},\Lambda_i^{\tau,e}$ ($N\times N$) are defined in the proof.
\end{lem}

The characterization of the equilibrium in Lemma \ref{lem:equil} is intuitive. Liquidations start from a benchmark $\overline\ell_i$ and increase (for positive $\Lambda_i^q$) in asset holdings while decreasing (for positive $\Lambda_i^\tau$) in the liquidation wedge. Higher liquidations result in a lower liquidation price (for positive $\Gamma$) due to more assets having to be absorbed by arbitrageurs (equation \ref{eqn:equil_gamma}). These means that liquidations by intermediary $i$ decrease in asset holdings and increase in the liquidation wedges applied to other intermediaries within the same sector and across different sectors (for positive matrix elements), a result of a substitution effect resulting from the increase in equilibrium price. In a model without fire sales, we have $\Lambda_j^{q,e}=\Lambda_j^{\tau,e}=0$. We distinguish in equation \ref{eqn:equil_ell} between the asset choices of intermediary $i$ and the wedges applied to intermediary $i$'s liquidations (denoted with no $\ast$) as opposed to equilibrium objects (denoted with $\ast$) that enter because they determine the equilibrium liquidation price.

\subsection{Regulator's Optimal Wedges}
We solve the regulator's problem by backward induction: first, we characterize the regulator's optimal intervention $\tau$ given a choice of model $M$ and signal $s$. We then characterize the regulator's optimal choice of model $M$. To simplify analysis, we assume that the regulator places equal weight on all intermediaries but places a welfare weight of zero on arbitrageurs.\footnote{This focuses attention on a distributive externality from shifting wealth between arbitrageurs and intermediaries. This can be incorporated by assuming that arbitrageurs have a high marginal value of wealth in the Beginning but cannot borrow, meaning that Pareto improvements are achieved by raising the liquidation price in the Middle and having a lump sum transfer from intermediaries to arbitrageurs in the Beginning (see e.g., \citealt{claytonschaab2025}).}

Given that liquidation wedges are revenue-neutral, the regulator's optimal choice of $\tau$ solves
$$\max_\tau \mathbb E[\sum_i U_i \, | \, s,M]-\delta(\tau),$$ 
subject to equilibrium determination (Lemma \ref{lem:equil}).

As preliminaries to the proposition below, we define $\overline\ell_i(q) = \overline{\ell}_{i}+\Lambda_{i}^{q}q_{i}-\sum_{i}\Lambda_{i}^{q,e}q_{i}$ to be the liquidations of intermediary $i$ if there is no regulatory intervention ($\tau=0$). We define $L(q)=\sum_i\overline\ell_i(q)$ to be total liquidations if there is no intervention. The following proposition characterizes optimal liquidation wedges in terms of these objects and model parameters.

\begin{prop}\label{prop:tau}
    Given a model $M$ and signal $s$, the regulator's optimal policy in the Middle is
    \begin{equation}\label{eqn:tau}
    \tau^\ast=\mathbb{E}\bigg[\Xi\bigg|s,M\bigg]^{-1}\mathbb{E}\bigg[(\sum_{i}\overline{\Lambda}_{i}^{\tau})^{T}\Gamma L(q)\bigg|s,M\bigg]
    \end{equation}
    where $\Xi=\overline{\Lambda}^{\tau T}+(\sum_{i}\overline{\Lambda}_{i}^{\tau})^{T}\Gamma(\sum_{i}\overline{\Lambda}_{i}^{\tau})+\Delta $, where $\overline\Lambda_i^\tau = (e_i\otimes I_N)^T\Lambda_i^\tau - (\Lambda_1^{\tau,e},\ldots,\Lambda_I^{\tau,e})$ is $N\times NI$ (where $e_i$ is the standard basis vector whose $i^{th}$ element is $1$ and $\otimes$ is the Kronecker product), and where $\overline\Lambda^\tau = (\overline\Lambda_1^{\tau T},\ldots,\overline\Lambda_I^{\tau T})^T$ is $NI\times NI$.
\end{prop}

The optimal wedges of Proposition \ref{prop:tau} are familiar from the macroprudential policy literature on fire sales, and intuitively encode an expected marginal cost-marginal benefit trade-off. The marginal cost of the policy is captured in the (conditional expectation of the) inverse matrix $\Xi$, while the marginal benefit is captured in the expectation.

There are three components to the private and regulatory marginal cost of liquidations, reflecting the distortion of intermediaries' activities away from their private optimum. First, increasing liquidation wedges directly distorts the intermediaries' activities (the first term of $\Xi$). Second, by using wedges to alter equilibrium prices, the regulator changes the incentives for intermediaries to sell different assets (the second term). Finally, there is the regulatory cost $\Delta$.

The social marginal benefit of regulation arises from the mitigation of the fire sale. This has two components that are central to our analysis. First is the causal effect of the policy intervention on equilibrium liquidation prices, captured by the term $(\sum_{i}\overline{\Lambda}_{i}^{\tau})^{T}\Gamma$. The utility consequence of these price changes is proportional to the total value of assets being sold by intermediaries, $L(q)$.

Importantly, this means the social benefit of regulation derives from a productive of the causal impact of the policy intervention, $(\sum_{i}\overline{\Lambda}_{i}^{\tau})^{T}\Gamma^{T}$, and the social value of changing the price, here given by $L(q)$. This means that in designing regulation, there is value not only to identifying causal policy impacts, but also to predicting the magnitude of liquidations $L(q)$. To fully disentangle these two mechanisms, we can assume independence of the predictive and causal components of the system.\footnote{Naturally absent this assumption, we could perform an analogous decomposition to Corollary \ref{cor:predictive}, but also include the covariance between the two terms.}

\begin{defn}[Predictive-Causal Independence]\label{defn:independence}
    We have predictive-causal independence if $\{\Lambda_i^\tau\}$ and $\{\overline\ell_i,\Lambda_i^q,\Lambda_i^{q,e}\}$ are independent of one another.
\end{defn}

Under predictive-causal independence, we obtain the following trivial corollary to Proposition \ref{prop:tau}
\begin{cor}\label{cor:predictive}
    Under predictive-causal independence, the regulator's optimal wedges are
    \begin{equation}
    \tau^\ast=\mathbb{E}\bigg[\Xi\bigg|s,M\bigg]^{-1}\underbrace{\mathbb{E}\bigg[(\sum_{i}\overline{\Lambda}_{i}^{\tau})^{T}\Gamma\bigg|s,M\bigg]}_{\textnormal{Causal: Policy Impact on Liquidation Price}}\overbrace{\mathbb{E}\bigg[L(q)\bigg|s,M\bigg]}^{\textnormal{Predictive: Total Liquidations}}
    \end{equation}
\end{cor}

An important implication of Proposition \ref{prop:tau} and Corollary \ref{cor:predictive} is that a model that is purely predictive -- that is, that can predict liquidations $L(q)$ -- can be valuable to a regulator, provided that the regulator has some prior or posterior knowledge over the causal impact of the policy. The predictive information informs the regulator as to the magnitude and margins for intervention by informing the regulator about the social benefit of the intervention. Corollary \ref{cor:predictive} implies that, holding fixed the causal impact of the policy, the regulator would design larger-magnitude interventions when the regulator's model predicted that total liquidations would be larger.

\subsection{Expected Welfare from a Model and Optimal Model Choice}
We next ask how a regulator in the Middle would choose a model $M\in\mathcal M$ under discretion, taking as given the asset holdings $q$ of intermediaries. Because choice of model in turn impacts intermediaries' optimal portfolio allocations (see Section \ref{sec:private_q}), we later consider model choice under commitment.

We begin by characterizing the expected utility to the regulator from a choice $M$ of model, and then characterize optimal model choice. As a preliminary to the proposition below, we define $\theta_i(q) = R_i - (\overline\gamma - \Gamma L(q))$ to be the fire sale losses to intermediary $i$ from selling assets prior to maturity in the event of no regulatory intervention. For the presentation in the main text, we focus on the case assuming predictive-causal independence (Definition \ref{defn:independence}). The proof of Proposition \ref{prop:model_welfare} characterizes the general case.
\begin{prop}\label{prop:model_welfare}
Under predictive-causal independence, the regulator's ex-ante expected welfare given positions $q$ and a model $M$ is
\begin{align}\label{eqn:model_welfare}
V(q,M)=&\mathbb{E}\bigg[q^{T}(R-p)-\frac{1}{2}q^{T}H^{q}q-\ell(q)^{T}\theta(q)-\frac{1}{2}\ell(q)^{T}H^{\ell}\ell(q)\bigg|s,M\bigg]
\\
&+\frac{1}{2}\mathbb{E}_{0}[\tau^{\ast T}]\Psi_{0}\mathbb{E}_{0}[\tau^\ast]+\frac{1}{2}\textnormal{tr}\bigg(\Psi_{0}\textnormal{cov}_{0}(\tau^\ast)\bigg)\nonumber
\end{align}
where $\Psi_0=\mathbb{E}_{0}\bigg[2(\sum_{i}\overline{\Lambda}_{i}^{\tau})^{T}\Gamma(\sum_{i}\overline{\Lambda}_{i}^{\tau})+\Delta+\overline{\Lambda}^{\tau T}H^{\ell}\overline{\Lambda}^{\tau}\bigg]$.
\end{prop}

Proposition \ref{prop:model_welfare} shows that the regulator's value function over assets $q$ and the model $M$ depends on two sets of terms.

The first line is the regulator's baseline welfare in the absence of intervention. Baseline welfare starts from the return on assets, $q^T(R-p)$, net of losses on assets sold prior to maturity, $-\ell(q)^T\theta(q)$. It then nets out the adjustment costs from the initial portfolio and from liquidations in the Middle. All of these terms are evaluated assuming no intervention, that is $\tau=0$. As a result, they do not depend on the regulator's choice of model $M$.

The second line is the welfare gains resulting from the regulator's optimally chosen intervention, which comprises two terms. The first term reflects the expected magnitude of the regulator's intervention, and so depends on $\mathbb E_0[\tau^\ast]$. Because interventions are targeted to the social benefit of intervening (equation \ref{eqn:tau}), this term is quadratic in the expected intervention. It is weighted by the consequences of intervention, reflected by the extent to which interventions move prices (the first term in $\Psi_0$), the regulatory costs (the second term), and the movement in holding costs from liquidations.

The welfare also depends on the accuracy of the regulator's model and intervention, captured in the second term that depends on the covariance matrix of the policy intervention, $cov_0(\tau^\ast)$. The intuition comes from the law of total variance: a perfectly informed regulator would target an intervention based on the true parameters of the system, $\tau^\ast = \Xi^{-1}(\sum_i\overline\Lambda_i^\tau)^T\Gamma L(q)$. A regulator that learned no information relative to the prior would have no covariance, so that all benefits would be reflected in the prior beliefs that are used to design the intervention.

In general, both of these terms depend on the choice of model $M$. The latter term depends on the choice of model because the choice of model determines the covariance matrix of the policy intervention. The former term also depends on the model, because the anticipated choice of model can affect the expected intervention. In particular, even under predictive-causal independence, from Proposition \ref{prop:tau} a model that informs about the causal impact of the policy intervention $\tau$ will inform about both the costs $\Xi$ of regulation and also about the causal impact $(\sum_i\overline\Lambda_i^\tau)^T$ of the policy intervention. We define a \textit{predictive} model as one that is uninformative about the causal structure of the policy intervention.

\begin{defn}\label{defn:predictive}
We say a model $M$ is \textnormal{predictive} if $\mathbb E[\Xi|s,M]=\mathbb E_0[\Xi]$ and $\mathbb E[\sum_i\overline\Lambda_i^{\tau T}|s,M]=\mathbb E_0[\sum_i\overline\Lambda_i^{\tau T}]$ for all signals $s$.
\end{defn}

\subsection{Optimal Predictive Model Choice}
We can now characterize the optimal choice of a model. We maintain simplicity by assuming predictive-causal independence for the main text (Definition \ref{defn:independence}) and by focusing on predictive models (Definition \ref{defn:predictive}).

We assume that the regulator faces a separable utility cost $\mathcal C(M,q)$ from adopting (predictive) model $M\in\mathcal M$. By Proposition \ref{prop:model_welfare}, for a predictive model (Definition \ref{defn:predictive}) the key sufficient statistic of the model from a welfare perspective is the prior covariance matrix over the ex-post policy intervention, $\Sigma_0^\tau = cov_0(\tau^\ast)$. We can therefore re-represent costs as $C(\Sigma_0^\tau,q) = \inf_{M\in \mathcal M | cov_0(\tau^\ast) = \Sigma_0^\tau}\mathcal C(M,q)$. We assume that $C$ is differentiable in $(\Sigma_0^\tau,q)$ over an open ball that contains its optimal value.

As a result, for a predictive model we can write the regulator's optimization problem as
$$\max_{\Sigma_{0}^{\tau}}\frac{1}{2}\textnormal{tr}\bigg(\Psi_{0}\Sigma_{0}^{\tau}\bigg)-C(\Sigma_{0}^{\tau},q).$$
We obtain the following result on the optimal predictive model.
\begin{prop}\label{prop:Mstar}
The regulator's optimal predictive model solves
\begin{equation}\label{eqn:Mstar}
\frac{\partial C(\Sigma_{0}^{\tau},q)}{\partial\Sigma_{0}^{\tau}}+\bigg(\frac{\partial C(\Sigma_{0}^{\tau},q)}{\partial\Sigma_{0}^{\tau}}\bigg)^{T}=\Psi_0
\end{equation}
where $ \frac{\partial C(\Sigma_{0}^{\tau},q)}{\partial\Sigma_{0}^{\tau}}$ is a square matrix whose ij-th element is $\frac{\partial C(\Sigma_{0}^{\tau},q)}{\partial(\Sigma_{0}^{\tau})_{ij}}$.
\end{prop}

Proposition \ref{prop:Mstar} yields an intuitive trade-off on optimal predictive model choice in terms of choice of covariance matrix of the policy intervention. The regulator is willing to pay a higher marginal cost to increase precision on dimensions where $\Psi_0$ is larger, that is when the policy has greater impact on aggregate liquidations and prices, when the regulatory cost is higher, or when the impact through holding costs is higher. In this sense, there is a complementarity between predictive power and knowledge of the causal policy impact: the regulator is willing to pay larger costs to acquire precise predictions of aggregate liquidations precisely on dimensions on which the policy impact on liquidations and liquidation prices is anticipated to be largest.

\subsection{Privately Optimal Asset Allocations}\label{sec:private_q}
We now turn to studying how model choice and ex-post intervention shape the ex-ante asset allocations of intermediaries. We begin by studying the private optimum of individual intermediaries, and then study socially optimal interventions.

Intermediary $i$ in the Beginning takes as given the model choice $M^\ast$ of the regulator and the resulting possible equilibria, and solves
$$\max_{q_{i}}\mathbb{E}_{0}\bigg[q_{i}^{T}(R_{i}-p_{i})-\ell_{i}^{T}(R_{i}+\tau_{i}^{\ast}-\gamma)-\frac{1}{2}q_{i}^{T}H_{i}^{q}q_{i}-\frac{1}{2}\ell_{i}^{T}H_{i}^{\ell}\ell_{i}\bigg].$$
Intermediary $i$'s asset allocation is therefore affected both through the specific intervention $\tau_i^\ast$ anticipated, and also through the liquidation price (which in turn also affects equilibrium liquidations). Intermediary $i$ knows that equilibrium liquidations are determined as in Lemma \ref{lem:equil}, but only internalizes the effect of its own $q_i$ on its own liquidations (and not on equilibrium liquidation price or equilibrium liquidations, and not on the optimal model choice or intervention).

We next turn to studying the effect of the regulator's model choice on ex-ante asset allocations of intermediaries. Note that Proposition \ref{prop:private_q} does not rely on predictive-causal independence or a predictive model.

\begin{prop}\label{prop:private_q}
The privately optimal asset allocation satisfies
\begin{align}\label{eqn:private_q}
\mathbb{E}_{0}\bigg[H_{i}^{q}q_{i}^{\ast}+\Lambda_{i}^{qT}H_{i}^{\ell}\overline{\ell}_{i}(q^{\ast})\bigg]=\mathbb{E}_{0}\bigg[R_{i}-p_{i}-\Lambda_{i}^{qT}\theta_{i}(q^{\ast})\bigg]-\mathbb{E}_{0}\bigg[\Lambda_{i}^{qT}\bigg(\tau_{i}^{\ast}-\bigg(H_{i}^{\ell}\overline{\Lambda}_{i}^{\tau}+\Gamma(\sum_{i}\overline{\Lambda}_{i}^{\tau})\bigg)\tau^{\ast}\bigg]
\end{align}
\end{prop}

Proposition \ref{prop:private_q} expresses the optimal portfolio choice in the form of a marginal cost-marginal benefit trade-off. The left hand side captures the marginal cost of increasing holdings of an asset, which includes both the ex-ante and ex-post adjustment costs of holding more and liquidating more of an asset. This term is evaluated at the no-intervention benchmark, that is as-if we had $\tau^\ast=0$.

The right hand side captures the marginal benefit, which is decomposed into a marginal benefit absent the ex-post regulatory intervention (the first term) plus the marginal benefit arising from the impact of intervention (the second term). The first term on the RHS is the baseline expected asset return, $R_i-p_i$, net of costs of liquidations. The liquidation costs are the amount liquidations are changed by increasing holdings, $\Lambda_i^{qT}$, times the fire sale loss in liquidation, $\theta_i(q^\ast)$.

The final term on the RHS captures the impacts of model choice and the ex-post intervention. This is the only term in equation \ref{eqn:private_q} that depends on model choice and intervention. It captures the cost of holding an asset induced through regulation. The expectation of ex-post policy intervention has two impacts. First, a higher wedge $\tau_{in}$ on intermediary $i$ increases the cost of liquidating asset $n$, discouraging the intermediary from holding portfolios that result in it liquidating that asset ex post. There are also equilibrium costs of the vector of wedges $\tau$ through the equilibrium price. In contrast, here a higher wedge $\tau_{in}$ encourages the intermediary $i$ to hold more of an asset when the asset price rises as a result, which partially offsets the benefit from raising the price. This is a standard channel of moral hazard.

It is clear that the ex-post intervention affects the optimal asset allocation: a higher expected tax on asset $n$ ex post directly discourages its purchase ex ante, but a higher liquidation price ex post encourages its purchase ex ante. As a result, even a purely predictive model can impact the asset allocation ex ante. In particular, even though under a purely predictive model (Definition \ref{defn:predictive}) the expectation of $\tau^\ast$ is that same as if the regulator ran no model, there is a covariance induced between the tax itself, $\tau^\ast$, and the impact on liquidations, $\Lambda_i^q$, as long as the predictive model is loading at least some on Bayesian inference on the impact of asset holdings on liquidations $\Lambda_i^q$. That is to say, focusing on the direct term $\mathbb E_0[\Lambda_i^{qT}\tau_i^\ast]$, for a purely predictive model we can write
$$\mathbb{E}_{0}\bigg[\Lambda_{i}^{qT}\tau_{i}^{\ast}\bigg]=\mathbb{E}_{0}\bigg[\Lambda_{i}^{qT}\bigg]\mathbb{E}_{0}\bigg[\tau_{i}\bigg]+\text{cov}_{0}\bigg(\Lambda_{i}^{qT},\mathbb{E}_{0}[\Xi]^{-1}\mathbb{E}_{0}\bigg[(\sum_{i}\overline{\Lambda}_{i}^{\tau})^{T}\Gamma\bigg]\mathbb{E}[L(q)|s,M]\bigg),$$
where the second term reflects how the true value of $\Lambda_i^q$ varies with the prediction of liquidations. In a similar fashion, moral hazard can also be exacerbated if the impact of asset allocations on liquidations $\Lambda_i^{q}$ is what a key driver of the predictability of ex post liquidations.

\subsection{Socially Optimal Asset Allocation}
Finally, we study the constrained efficient asset allocation $q$ of the regulator. Formally, we assume that the regulator can impose a wedge on asset holdings, that is a revenue-neutral tax $t_i=(t_{i1},\ldots,t_{iN})^T$. Because the regulator has a complete set of wedges ex ante, this is equivalent to the regulator directly picking portfolio allocations ex ante.\footnote{It is straight-forward to extend results to incomplete ex-ante instruments, but for brevity we focus on complete instruments.} The following proposition characterizes the optimal wedges $t=(t_1^T,\ldots,t_I^T)^T$ when the regulator chooses an optimal predictive model ex post.

\begin{prop}\label{prop:ce_q}
For a predictive model, the social planer's optimal ex-ante portfolio wedges $t$ are
\begin{align}\label{eqn:ce_q}
t=&\mathbb{E}_{0}\bigg[(\sum_{j}\overline{\Lambda}_{j}^{q})^{T}\Gamma L(q)-\Lambda^{q,eT}\sum_{i}\bigg(\theta_{i}(q)+H_{i}^{\ell}\ell_{i}(q)\bigg)\bigg]\nonumber
\\
&+\frac{dC(\Sigma_{0}^{\tau},q)}{dq}-\mathbb{E}_{0}\bigg[(\sum_{i}\overline{\Lambda}_{i}^{q})^{T}\bigg]\Upsilon_{0}^{T}\Psi_{0}\mathbb{E}_{0}[\tau^{\ast}]-\mathbb{E}_{0}\bigg[\Psi\tau^\ast\bigg]
\end{align}
where $\Upsilon_{0}=\mathbb{E}_{0}\bigg[\Xi\bigg]^{-1}\mathbb{E}_{0}\bigg[(\sum_{i}\overline{\Lambda}_{i}^{\tau})^{T}\Gamma\bigg]$ and where $\mathbb E_0[\Psi\tau]$ is a stacked vector whose $i^{th}$ block is $\mathbb{E}_{0}\bigg[\Lambda_{i}^{qT}\bigg(\tau_{i}-\bigg(\Gamma(\sum_{i}\overline{\Lambda}_{i}^{\tau})+H_{i}^{\ell}\overline{\Lambda}_{i}^{\tau}\bigg)\tau\bigg)\bigg]$.
\end{prop}

The optimal ex-ante asset tax formula is intuitive and is decomposed into two lines. The first line captures the uninternalized effects of intermediaries increasing holdings of an asset in absence of regulatory intervention. First, as $q$ changes, total liquidations change across intermediaries, resulting in changes in the liquidation price and hence changes in revenue proceeds from total liquidations $L(q)$. Second, changes in forced liquidations through changes in the equilibrium price result in losses due to the size of the discount $\theta_i(q)$ and due to changes in marginal holding costs $H_i^\ell \ell_i(q)$. In absence of an ex-post intervention, this first line would capture the standard macroprudential regulation of asset positions.\footnote{Since our framework allows negative positions $q_{in}<0$, we can think of these negative positions as liabilities. As such, it also captures regulation of such liabilities.}

The second line captures the interaction of the ex-ante asset regulation with the ex-post model design and intervention. The first term, $dC/dq$, reflects how the changing assets held by intermediaries affects the cost of acquiring information through the model ex post. The regulator imposes larger holding taxes on asset positions that make running the model more costly ex post. For example, the regulator might discourage holdings of non-transparent or hard-to-assess asssets.

The second term on the second line captures the effect on the expected intervention size (the term $\mathbb E_0[\tau^{\ast T}]\Psi_0\mathbb E_0[\tau^\ast]$ in equation \ref{eqn:model_welfare}). Intuitively, concentrations into a position $q$ increase total liquidations on the margin by $\sum_i\overline\Lambda_i^q$, which leads the regulator ex post to increase the size of the ex-post intervention $\tau^\ast$ accordingly. This increase in ex-post intervention helps to manage the fire sale and so mitigates the impact of the ex-ante increase in asset allocation. This gives a first dimension whereby the ex-post intervention provides a partial substitute for the ex-ante asset tax, and leads the regulator to require potentially smaller interventions ex ante, knowing that fire sales will be managed ex post.

Finally, there are the direct and moral hazard effects on intermediaries, $\mathbb E_0[\Psi\tau^\ast]$, are those in the final term in equation \ref{eqn:private_q} of Proposition \ref{prop:private_q}. First, the prospect of the ex-post tax on liquidations directly discourages holdings of the asset ex ante when holding more promotes liquidations ex post. Second, there is the moral hazard effect: as the liquidation price rises, intermediaries' perceived cost of liquidations $\theta_i(q)$ falls, and so they become more willing to hold more of an asset even if that forces them to liquidate. It is interesting again to observe that for assets for which the direct effect dominates the indirect price effect, the ex-post tax actually serves as a substitute for ex-ante regulation, leading to a smaller intervention $t$ ex ante.

\subsection{Ex-Post Subsidy-Based Interventions}\label{sec:subsidy}
Our baseline analysis assumes that the regulator directly manages liquidations through a wedge/tax on liquidations. In practice, ex-post interventions can also involve ``bailout'' interventions that subsidize retaining assets rather than taxing selling them (e.g. lender of last resort, debt guarantees). Suppose that rather than taxing liquidations $\ell_i$, the regulator instead applies a subsidy on the amount $q_i-\ell_i$ of the asset that is held to maturity. Formally, the payoff to intermediary $i$ in the End is now
$$U_i = U_i + (q_i^T-q_i^{\ast T})\tau_i - (\ell_i^T - \ell_i^{\ast T})\tau_i,$$
so that the regulator continues to apply revenue-neutral interventions.\footnote{Since individual intermediaries take revenue remissions as given, there is still moral hazard from the subsidy.}

Because the regulator in the Middle takes the asset allocations as given, the regulator's optimization problem over both model choice and the ex-post intervention are formally the same as before. Thus, Lemma \ref{lem:equil} and Propsitions \ref{prop:tau}-\ref{prop:Mstar} continue to apply. However, the privately optimal asset allocation is affected by the subsidy on asset retention. In particular, in Proposition \ref{prop:private_q}, the expected return on holding asset $q_i$ is raised from $R_i$ to $R_i+\tau_i^\ast$. Intuitively an ex-post subsidy promotes overinvestment in that asset. This gives rise to a familiar channel of moral hazard.

Relative to the case of a liquidation tax, it is worthwhile to note that the only new term for private intermediaries in their optimization problem is the revenue benefits $q_i^T \mathbb E_0[\tau_i^\ast]$ of their asset position. Therefore, relative to the liquidation tax, the regulator's model choice only affects intermediaries' asset allocation ex ante to the extent it changes the expected size of the intervention, $\mathbb E_0[\tau_i^\ast]$. In particular for a predictive model under predictive-causal independence (Definitions \ref{defn:independence} and \ref{defn:predictive}), we know that $\mathbb E_0[\tau_i^\ast]$ does not depend on the model used; instead, it is only the covariance matrix of the intervention (its precision) that deepends on the model. It is thus interesting and surprising that purely predictive models that help with prediction but not with causal inference do not exacerbate moral hazard relative to the case of the liquidation tax. In contrast, models that are informative about the causal impact of the policy intervention change expectations of the policy intervention, and are potentially associated with moral hazard. The logic is reminiscent of that of \cite{laffont1986} in the context of regulation of a firm with unobservable effort and uncertain costs.

\subsection{Extensions}

\paragraph{Better Informed Intermediaries.} Our model assumes that in the Beginning, intermediaries and regulators share a common prior over the model parameters. In practice, intermediaries may be better informed about model parameters. One could extend the model by assuming that intermediaries had a more precise prior than regulators. This would lead a regulator to also infer information about the structural parameters of the economy from observing portfolio holdings directly. One possible method of incorporating is to interpret $C(M,q)$ as a cost of the inference the regulator is undertaking, and so interpret the model as both a processing of information (including from such Bayesian inference). One possible advantage of the ex-post intervention would be its ability to react to information discovered from observing intermediaries' choices, which might reduce risks of moral hazard or imperfectly calibrated regulation ex ante. That is, better-informed intermediaries who saw a regulator was under-regulating an asset $n$ ex ante would also know that the regulator would discover the mistake ex post, and so intervene more strongly upon it. Exploring the effects of differential information between the regulator and intermediaries is an interesting direction.

\paragraph{Dynamic Learning.} We embed a one-shot learning problem. One could extend our framework to embed dynamic information acquisition by assuming that our baseline Beginning-Middle-End model was a stage game played at each date $t=0,1,\ldots$. In this environment, one could think of there being a true distribution $\mu^\ast$ from which model parameters are drawn each period, so that the regulator and intermediaries learn about this true distribution over time. The regulator and intermediaries would carry information forward at each date, and so the regulator would consider both how a model acquired information on the current crisis and also how it informed about the underlying variables. We conjecture that this would make predictive models relatively myopically useful: they would give potentially substantial information for intervening in the current crisis, but relatively little information about the underlying structural parameters, and so might be of limited use in updating beliefs about the true distribution of parameters. This could face the regulator with an interesting dynamic trade-off between myopically acquiring more predictive information, and trying to uncover the true structural parameters that would also be useful for designing regulation and interventions during the next crisis.

\paragraph{Commitment vs.\ Discretion in Model Choice.} We have assumed the regulator chooses the model and ex-post policy intervention with discretion, after asset allocations are chosen. As highlighted by the results in Section \ref{sec:private_q}, this leads to a potential time consistency problem in model choice since the model choice affects ex-ante asset allocations. This time consistency problem will likely be more pronounced the more incomplete the regulator's ability to regulate portfolio positions is ex ante, for example if the regulator cannot regulate all intermediaries or all assets. In our ongoing work, we are exploring the implications of commitment versus discretion for optimal model design and welfare gains from use of a predictive model.

\section{Graph Representation Learning for Holdings Data}
\label{sec:graph}

We now move to an empirical examination of whether high-dimensional forecasting models can in fact be used successfully for macroprudential applications. This analysis establishes several points. First, it provides a blueprint for the practical implementation of deep learning models by central banks and other financial supervisors. Second, it demonstrates that such models do in fact have significant predictive power---which is a crucial motivating fact for the theory. Third, motivated by the theoretical setting, it lays out several key design principles that deep learning architectures should follow when applied to financial holdings data.

\paragraph{GNN Architectures and Holdings Data.} We begin by introducing a deep learning architecture tailored to holdings data, and we discuss why it is optimally suited to this setting. Much of the modern deep learning toolkit is optimized for grid inputs such as images and sequence inputs such as text. Indeed, architectures that have proved successful are those that exploit the particular structure of the data they are modeling: examples include convolutional neural networks for images and other grid-structured data, and text transformers---which underpin modern large language models---for textual data. Financial holdings data does not fit neatly into either of these categories. Instead, its defining feature is that it has very rich \textit{relational structure}: the data can be naturally thought of as representing a graph connecting investors and assets, with the information relevant to the learning task contained in the graph's edges, i.e.\ the positions connecting investors to assets.  

Graph neural networks (GNNs) are a class of deep learning models that specifically models and exploits relational structure in the data (\citealt{scarselli2008graph}; \citealt{hamilton2017inductive}; \citealt{wu2020comprehensive}). By iteratively propagating and aggregating information along the edges that connect graph nodes (in this case, assets and investors), GNNs learn embeddings for each node in the graph---i.e., representations of the nodes in a latent vector space that effectively capture the characteristics relevant to the tasks the model is trained against. In practice, in our implementation an investor’s embedding evolves based on the embeddings of the assets they hold, and those asset embeddings, in turn, adjust in light of the embeddings of the investors that include them. Through iterated rounds of this neighbor‐aggregation mechanism (also known as \textit{message passing}), information flows along the network of positions, allowing the model to learn explicitly from the relational structure of the holdings data.

GNNs have state-of-the-art empirical success in domains where relational structure is paramount. The protein-folding model AlphaFold, for example, uses a graph-based architecture to encode the three-dimensional interactions among amino acids, and it has dramatically advanced the field of protein structure prediction (\citealt{jumper2021highly}). Traffic-forecasting systems used for products such as Google Maps represent road networks as graphs and learn congestion and travel-time patterns directly from the topology of streets and highways (\citealt{derrow2021eta}). Similarly, in frontier drug discovery models, molecules are treated as graphs of atoms and bonds, and GNNs have revolutionized the prediction of chemical properties and binding affinities (\citealt{jiang2021could}). Each of these breakthroughs rests on the fundamental ability of GNNs to natively handle and learn from graph-structured data.

In the context of asset holdings data, GNNs have two key properties that other deep architectures lack: permutation invariance and inductive learning. First, the models are \textit{permutation invariant}, meaning that the learning process and estimation results do not depend on any arbitrary relabeling or reshuffling of investor and asset identifiers. The architecture attains permutation invariance because aggregation operates over unordered sets of neighbors, and by doing so they respect a fundamental symmetry of the problem. 

Second, the GNN architecture features \textit{inductive learning}. Crucially, all of the model's parameters are shared, in the sense that there are no parameters specific to particular assets or investors. Therefore the same learned representation rules apply across all nodes and edges: given a graph $\mathcal{G}$, regardless of the number and identity of the nodes, the trained GNN architecture is able to construct asset and investor embeddings simply from the graph structure and the node characteristics. Inductive generalization thus follows naturally from parameter sharing: the model can immediately generate embeddings and forecasts for new, unseen investors or assets without any retraining. This feature is not only valuable for real-time regulatory applications, but also it enforces a strong form of regularization upon the model, preventing overfitting of the training data and leading to good generalization to unseen, out-of-sample data.

This combination of relational representation learning, proven empirical performance, permutation invariance, and inductive design makes GNNs a particularly well suited deep learning architecture for the holdings data setting and for macroprudential applications. Our specific implementation is a \textit{graph transformer}, which incorporates an attention mechanism in the basic GNN architecture, as we discuss below.

\paragraph{Core Architecture Specification.} To formally specify the GNN graph transformer architecture, we start by modeling the holdings data as a bipartite graph---meaning a graph with two distinct types of nodes, and whose edges only connect nodes of the two different types. The holdings data forms a bipartite graph in which investors connect to assets via position edges.  Formally, we let a given cross-section of the data be represented as $\mathcal G=(\mathcal I,\mathcal A,\mathcal E)$, where $\mathcal I=\{1,\dots,I\}$ indexes investors, $\mathcal A=\{1,\dots,N\}$ is the set of assets, and $\mathcal{E}$ is the set of edges (i.e., positions). A certain position exists in the graph if the corresponding investor-asset pair is in the set of edges: whenever investor $i$ holds asset $a$, letting $w_{ia}$ be the size of the position, we have that $(i,a,w_{ia})\in\mathcal E$.\footnote{The positions can alternatively be written as $w_{ai}=w_{ia}$. We allow for both notations so as to keep the rest of the formal architecture specification symmetric.} We write $\mathcal{V}=I\cup\mathcal A$ as the set of all nodes, both assets and investors. We let $N(v)$ be the neighborhood set for a given node $v$. This is the set of all nodes that are connected to $v$: for assets, $N(v)$ corresponds to the investors holding the asset, while for investors this corresponds to the set of assets in the investor's portfolio.

Each node $v\in\mathcal{V}$ has an associated set of characteristics $x_v$. The characteristics can be either numerical (e.g., the total amount outstanding of a given security) or categorical (e.g., the type of institutional investor). For categorical features, these are pre-embedded to a vector space of dimension $d_{c}$ using maps that are learned jointly with the rest of the model's parameters. The node characteristics vector $x_v$ concatenates over all individual features, both scalar-valued ones and vector-valued categorical ones.

The graph transformer learns node embeddings (both asset embeddings and investor embeddings) at various hierarchical levels of information aggregation. The first-level embeddings, which we denote as $h_v^{(0)}$, are obtained simply by embedding the node characteristics vectors $x_v$ into a $d_h$–dimensional hidden space via a learnable map $\phi$:
\begin{equation}
  h_v^{(0)} \;=\;\phi(x_v)\,.
\end{equation}
Next, the architecture passes these layer-zero node embeddings through several successive layers of message passing which aggregate information over the graph: this message passing stage is the core of the GNN model and is what allows the model to learn from the data's relational structure. To increase the expressivity and learning capability of the model, we integrate attention mechanisms in each of the GNN message passing layers, which allow the architecture to learn from the data which positions should be given more or less weight in any given information aggregation steps. This incorporation of attention layers is what makes our architecture a graph transformer.

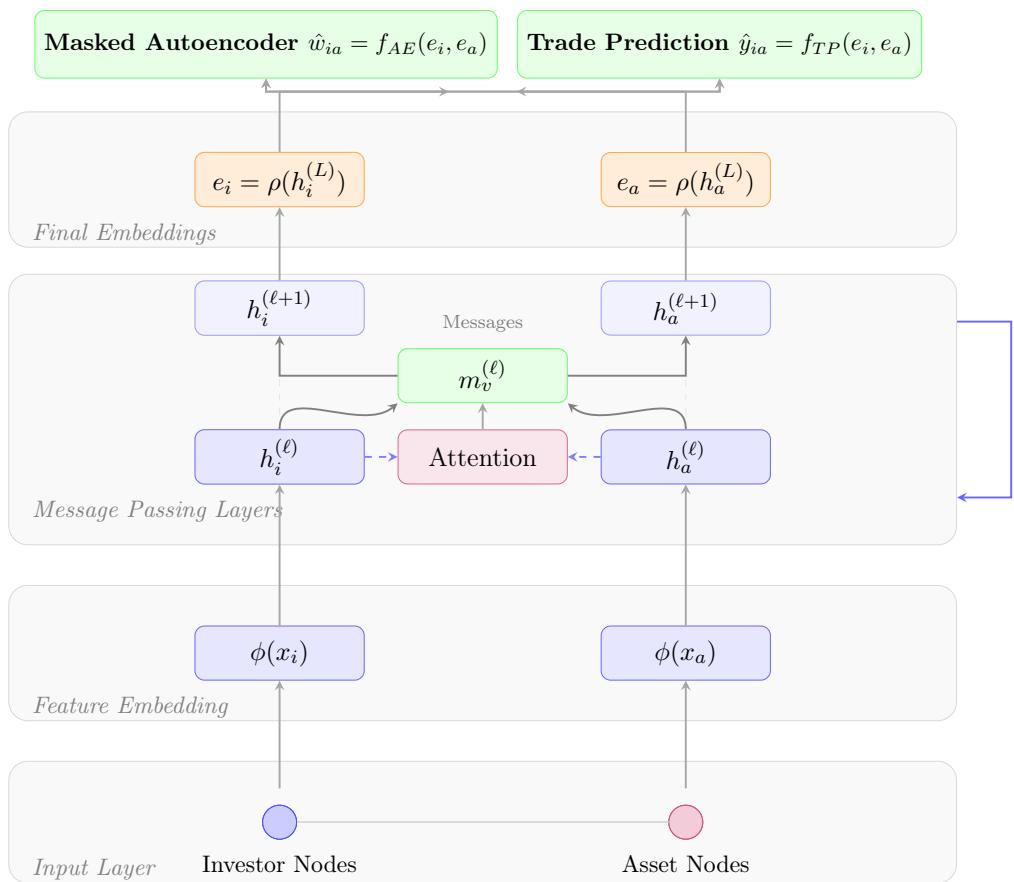
\begin{figure}[!htp]
    \centering
    \caption{\textbf{Model architecture} }
    \scalebox{.9}{
        \input{graphs/architecture_tikz}
    }
    \medskip
    
    \small \textit{Notes:} We visualize the architecture of our graph transformer model diagrammatically.
    \label{fig:model-architecture}
\end{figure}

We perform $L$ distinct layers of message passing. The message passing at layer $\ell$ unfolds in several distinct steps. First, the prior-layer node embeddings are passed through a feed-forward layer $M^{(\ell)}:\mathbb R^{d_h}\to\mathbb R^{d_h}$ to construct \textit{node messages}:\footnote{All feed-forward layers in our architecture use GELU non-linearities (\citealt{hendrycks2016gaussian}). These help avoid dead gradients during training.}
\begin{equation}
    M_{v}^{\ell} = M^{(\ell)}\bigl(h_v^{(\ell-1)}\bigr).
\end{equation}
Next, the attention mechanism computes aggregation weights that dictate how the individual node messages will be weighted when passing information over the graph. We allow for $S_A$ distinct and independent attention heads. Each attention head $s=1,\dots,S_A$ forms attention weights using learnable projection matrices $W_{qs},W_{ks}\in\mathbb R^{d_h\times d_h}$ which project prior-layer embeddings into query and key spaces, respectively---akin to how text transformers form attention values for the interactions of tokens in a text sequence (\citealt{vaswani2017attention}). The attention weight $\alpha_{vu}^{(\ell)}$ indicates the importance of messages from each node $u$ in the neighborhood $N(v)$ of node $v$ at layer $\ell$, and it is formed by averaging over the individual attention heads:
\begin{equation}
    \alpha_{vu}^{(\ell)}
    =\frac{1}{S_A}\sum_{s=1}^{S_A}\frac{\exp\bigl\{(W_{qs}\,h_v^{(\ell-1)})^T(W_{ks}\,h_u^{(\ell-1)})\bigr\}}
          {\sum_{u'\in\mathcal N(v)}\exp\bigl\{(W_{qs}\,h_v^{(\ell-1)})^\top(W_{ks}\,h_{u'}^{(\ell-1)})\bigr\}}.
\end{equation}
Having established attention weights $\alpha_{vu}^{(\ell)}$, the message passing algorithm then constructs an aggregated message $m_v^{(\ell)}$ for each node $v$ by averaging over the messages received from each of its neighbors $u\in\mathcal N(v)$, according to:
\begin{equation}
  m_v^{(\ell)}
    =\sum_{u\in\mathcal N(v)}
      w_{vu}\,\alpha_{vu}^{(\ell)}\,M_{u}^{(\ell)},
  \qquad
\end{equation}
The node embeddings at the next hierarchical stage $\ell$ are then updated using another feed-forward layer $U^{(\ell)}:\mathbb R^{2d_h}\to\mathbb R^{d_h}$ which maps the current embeddings and the current aggregated messages into the next-stage representations:
\begin{equation}
  h_v^{(\ell)}
    =U^{(\ell)}\bigl(h_v^{(\ell-1)},\,m_v^{(\ell)}\bigr).
\end{equation}
After $L$ rounds of message-passing, a readout feed-forward layer $\rho:\mathbb R^{d_h}\to\mathbb R^{d_e}$ produces final asset and investor embeddings:
\begin{equation}
  e_v = \rho\bigl(h_v^{(L)}\bigr).
\end{equation}
When referring to node embeddings in subsequent sections, unless otherwise specified we refer to these final representations $e_v$.  Because all parameters are shared across nodes and layers, this architecture is both permutation invariant and inductive, allowing predictions on unseen investors or assets without retraining.

\paragraph{Prediction Heads.} To train and leverage the learned embeddings $e_v$, we attach two task‐specific heads to the graph transformer. The first task uses a masked autoencoder (MAE) objective, where we randomly mask a subset of edges (including non-existent edges drawn at random, which we represent using $w_v=0$) and ask the model to predict whether the edge exists ($w_v\neq 0$) and the associated position size $w_v$. The second task uses a supervised trade prediction objective, where we ask the model to engage in a pure forecasting task, using embeddings computed using time $t$ holdings information to predict the cross-sectional pattern of investor trades in the future, between time $t$ and time $t+1$.

For the masked autoencoder (MAE) objective, we randomly mask a subset of edges and predict masked weights $\hat w_{ia}=f_{\mathrm{AE}}(e_i,e_a)$ using a feed-forward head layer $f_{\text{AE}}$ which takes as inputs the embeddings for the given masked investor-asset pair $(i,a)$. The autoencoder objective minimizes a mean squared error loss defined over the divergence between the true edge weights $w_v$ and the predicted edge weights $\hat{w}_v$:
\begin{equation}
  \mathcal L_{\mathrm{AE}}
    =\sum_{(i,a)\in\mathcal{V}_\text{masked}}
     \bigl(w_{ia}-\hat w_{ia}\bigr)^{2},
\end{equation}
where $\mathcal{V}_\text{masked}$ denotes the set of masked edges.

For trade prediction, we construct the targets by first defining the percent changes in holdings for a given position in asset $a$ by investor $i$ between time $t$ and time $t+1$ (in practice, quarters) as:
\begin{equation}
  \Delta\%\!q_{ia,t}
    =\frac{q_{ia,t}-q_{ia,t-1}}{q_{ia,t-1}}.
\end{equation}
We strip away all common movement in trades for a particular asset $a$, such as for example movement induced by changes in the asset's valuation which affect all investors. To do this, we construct cross-sectional z-scores of the trades $\Delta\%\!q_{ia,t}$, which subtract the average percentage position change for asset $a$ in the given time period ($\overline{\Delta\%\!q_{a,t}}$) and divide by the standard deviation of the same position changes ($\sigma(\Delta\%\!q_{a,t})$), placing all assets in all time periods on the same scale. The cross-sectional trade z-scores are thus defined as:
\begin{equation}
  y_{ia,t}
    =\frac{\Delta\%\!q_{ia,t}-\overline{\Delta\%\!q_{a,t}}}
           {\sigma(\Delta\%\!q_{a,t})}.
\end{equation}
The cross-sectional trade patterns captured by $y_{ia,t}$ are the targets for the supervised prediction head. Specifically, we construct predicted trades $\hat y_{ia}=f_{\mathrm{TP}}(e_i,e_a)$ using a feed-forward layer head $f_{\text{TP}}$ which acts on the relevant pair of asset and investor embeddings.\footnote{While for compactness we are not carrying through time subscripts on the embeddings $e_v$, naturally for the trade prediction task we use embeddings estimated using the graph $\mathcal{G}$ at time $t-1$ to construct the predictions for time $t$, $y_{ia,t}$.} The supervised objective is again defined over the mean squared error between the actual trades and the predicted trades:
\begin{equation}
  \mathcal L_{\mathrm{TP}}
    =\sum_{(i,a)\in\mathcal{V}}\bigl(y_{ia}-\hat y_{ia}\bigr)^{2}.
\end{equation}

\paragraph{Model Training.} To recap, the model contains several learnable components, all of which are parameterized using a high-dimensional set of parameters. The trainable components include the feature map $\phi$, the pre-embedding functions for categorical characteristics, the message feed-forward layers $M^{(\ell)}$, the attention mechanism projection matrices $W_{qs}$ and $W_{ks}$, the node update function $U^{(\ell)}$, the embeddings projection layer $\rho$, and the task-specific prediction heads $f_{\text{AE}}$ and $f_{\text{TP}}$. The model architecture is summarized visually in Figure \ref{fig:model-architecture}. We collect the set of parameters in all these learnable components in the vector $\Theta$.

The model is trained end-to-end by minimizing a joint loss which combines the mean squared error losses from the two training tasks:
\begin{equation}
  \min_{\Theta}\;
    \mathcal L(\Theta)
  =\mathcal L_{\mathrm{AE}}
   +\kappa\,\mathcal L_{\mathrm{TP}},
\end{equation}
where $\kappa>0$ determines the relative weight of the two training tasks. We optimize the model parameters $\Theta$ using the Adam optimizer (\citealt{kingma2014adam}).

\begin{table}[!htp]
  \centering
  \caption{\textbf{Model hyperparameters}}
  \label{tab:hyper}
  \begin{tabular}{lc}
    \toprule
    Hidden dimension ($d_h$)       & 256   \\
    Embedding dimension ($d_e$)    & 128   \\
    Layers ($L$)                   & 3     \\
    Attention heads                & 4     \\
    Dropout rate                   & 0.1   \\
    Learning rate                  & $10^{-2}$ \\
    Weight decay                   & $10^{-5}$ \\
    Loss weight ($\kappa$)         & 1     \\
    \bottomrule
  \end{tabular}
      \medskip\smallskip
    
    \small \textit{Notes:} We list the hyperparameters used for our graph transformer architecture and for training.
\end{table}

\paragraph{Optimality of Graph-Based Architectures for Holdings Data.} Before moving on the empirical implementation, we discuss more precisely the sense in which message-passing, graph-based architectures are optimal in the context of holdings-based problems. To do this, we lay out a few definitions. Let $W\in\mathbb{R}^{I\times N}$ be the full holdings matrix with entries $w_{ia}$, and let $f\colon\mathbb{R}^{I\times N}\to\mathbb{R}^{d}$ be a functional acting on the graph $\mathcal{G}$ represented by $W$. A continuous graph functional is permutation-invariant if, for all permutation matrices $P_1\in\mathbb{R}^{I\times I},\;
P_2\in\mathbb{R}^{N\times N}$, it satisfies $f(P_1 W P_2^T)=f(W)$. Informally, permutation invariance means that if we were to arbitrarily relabel columns and rows of the holdings matrix (i.e., investors and assets), the output of $f$ would remain the same. All regulatory or prediction targets we care about (future trading patterns, systemic risk scores, etc.) are assumed to lie in this family, reflecting the economics of the problem.

A well-known idea in the literature on graph deep learning is that in order to represent a permutation-invariant mapping with the fewest parameters, one should enforce permutation invariance via shared (message‑passing) parameters (\citealt{zaheer2017deep}, \citealt{maron2018invariant}, \citealt{xu2018powerful}). This is the sense in which the models are optimally sample-efficient. For illustration, compare two classes of models. First, consider the class of GNNs which implement permutation-invariant message-passing layers as described above, with shared parameters. Second, consider the class of sequence or grid networks that act on the flattened matrix $\operatorname{vec}(W)$ under a fixed but arbitrary ordering of rows and columns (this class includes recurrent neural networks, convolutional neural networks, sequence transformers, and so on). Intuitively, GNNs can approximate permutation-invariant graph functionals without carrying superfluous degrees of freedom that sequence/grid networks would have to ``use up'' to relearn permutation invariance from data. Message-passing GNNs are efficient because the architecture is itself permutation-invariant, avoiding the need to use additional parameters to learn and enforce it.

\section{Empirical Implementation}
\label{sec:empirical}

We train our deep learning model on quarterly institutional holdings from Factset, starting in 2005Q1. The training-validation split is done at the level of quarters: we sort quarters into training and validation sets. The holdings data gives us observations of the positions graph $\mathcal{G}$ for each quarter's cross-section, and we also use it to construct the standardized trade indicators $y_{ia,t}$. We do not include any quarters following 2019Q3 in the training set, so that we can use the Covid crisis of 2020 as a particularly strict test episode, in the sense that the Covid quarters are not just out of the training sample, but also the model is only trained with data prior to the start of the crisis, mimicking the way in which the model would be deployed in an actual regulatory scenario.

We construct the node feature vectors $x_v$ using reference information from Factset as well as from the Global Capital Allocation Project (GCAP) security master file (\citealt{coppola2021redrawing}, \citealt{coppola2025safe}). For assets, the feature vectors $x_v$ include asset class, currency of denomination, amount outstanding, number of holders, average position size, standard deviation of position size, as well as bond sub-class and coupon for debt securities. For investors, they include institution type (such as open-end mutual funds, ETFs, separate accounts, etc.), manager style (including flags for active vs.\ passive portfolio management and strategy types), total AUM, number of positions, average position size, and standard deviation of position size. In principle, our architecture also allows for the use of global features that vary over time but not across nodes: these can be introduced as vectors which enter message-passing in the same way for all nodes in a given time period. In ongoing work, we are integrating global features and assessing the impact on the model's performance.\footnote{The weight to be placed on global features can be made learnable by the model. Global features may include time series measures such as aggregate credit spreads and other macro series.} Similarly, we are exploring the use of price data both as predictive features and targets for the model.

\begin{figure}[!htp]
  \begin{center}
  \caption{\textbf{Performance metrics}}
  \label{fig:performance}
  \includegraphics[width=.7\textwidth]{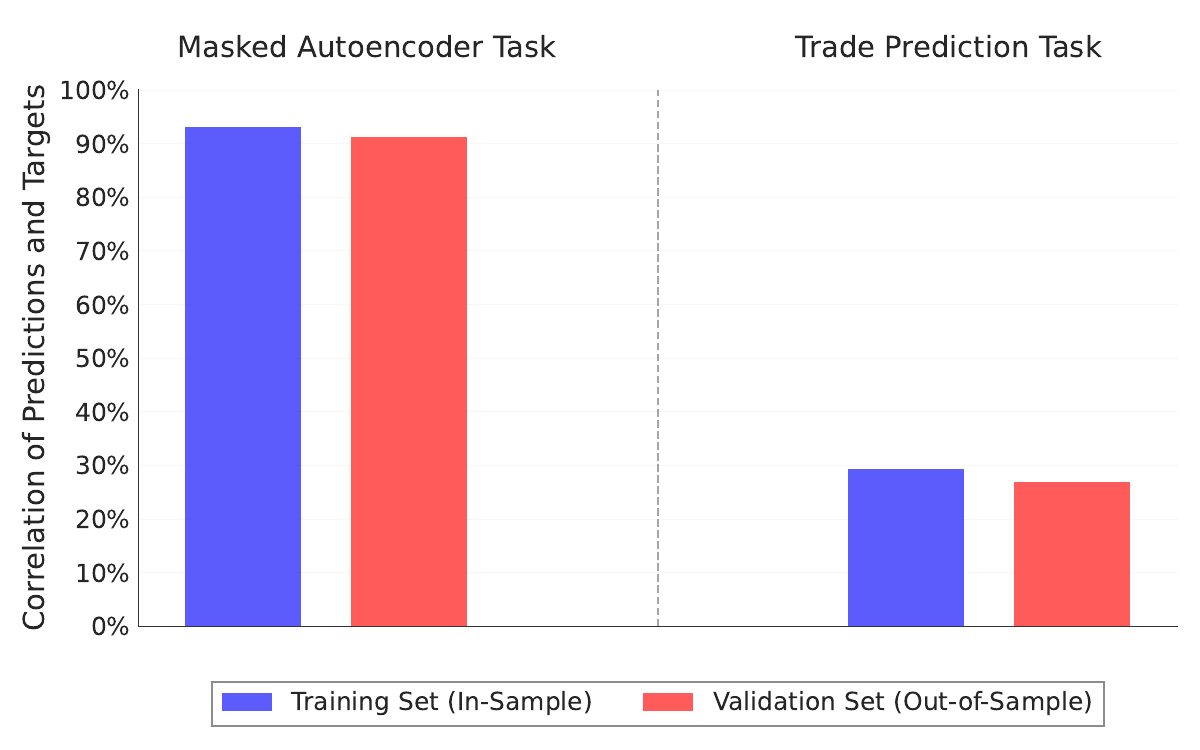}
  \end{center}
    \small \textit{Notes:} We plot the correlations between the trained model's predictions and the targets for the two tasks (masked autoencoder and cross-sectional trade prediction). The blue bars show performance on the training set, while the red bars show out-of-sample performance on the validation set.
\end{figure}

Hyperparameters are chosen as in Table \ref{tab:hyper}. In particular, we set the hidden dimension to $d_h=256$ and the final embeddings dimensionality to $d_e=128$. We use $L=3$ layers of message-passing: we do not increase $L$ beyond this number to prevent over-smoothing problem, whereby the node embeddings would converge to similar values: intuitively, over-smoothing occurs for higher numbers of message-passing iterations since each consecutive iteration increases the \textit{receptive field} of each node, i.e.\ the set of nodes that the final-layer embeddings attend to, and for high $L$ values the receptive fields of all nodes converge to the largest possible field, which is the set of all nodes in the graph. We allow for four attention heads and we give equal weight to the two training tasks by setting $\kappa=1$. Altogether, the model has a total of 3,640,465 parameters.

We introduce dropout during training for additional regularization, with a dropout rate of $0.1$. The Adam optimizers uses a starting learning rate of $10^{-2}$ with progressive weight decay. We train the model using a compute node with four NVIDIA H100 GPUs. We train for up to 500 epochs, with early stopping based on the loss on the validation sample.

Out‐of‐sample, the MAE head achieves a correlation above 0.90 between reconstructed and true positions, indicating that the GNN captures structural regularities in holdings. As mentioned in the introduction, the autoencoder prediction is best interpreted in the context of the model's parameter-to-data ratio: in this case, the model's roughly 3.6 million parameters represent less than 1\% of possible investor-asset pairs in the data. The trade‐prediction head yields an average correlation of just under 0.30 between predicted and realized trade indicators, with minimal degradation from training to validation sets.  Figure~\ref{fig:performance} plots the correlation between the trained model's predictions and the targets for both tasks: the blue bars show the performance on the training sample, while the red bars show the out-of-sample performance on the validation set. The fully inductive design, sharing parameters across nodes and layers, ensures stable performance on unseen investors and assets, such that the model performs very similarly out-of-sample as it does on the training data. In ongoing work, we are performing a descriptive analysis of the asset and investor embeddings produced by the model, so as to provide greater interpretability of the model's predictions.

\begin{figure}[!htp]
\begin{center}
\caption{\textbf{Performance metrics: heterogeneity}}\label{fig:performance-het}
    \begin{subfigure}{0.65\textwidth}
        \centering
        \caption{Validation set performance, by subsample}\label{fig:het-a}
        \includegraphics[width=\textwidth]{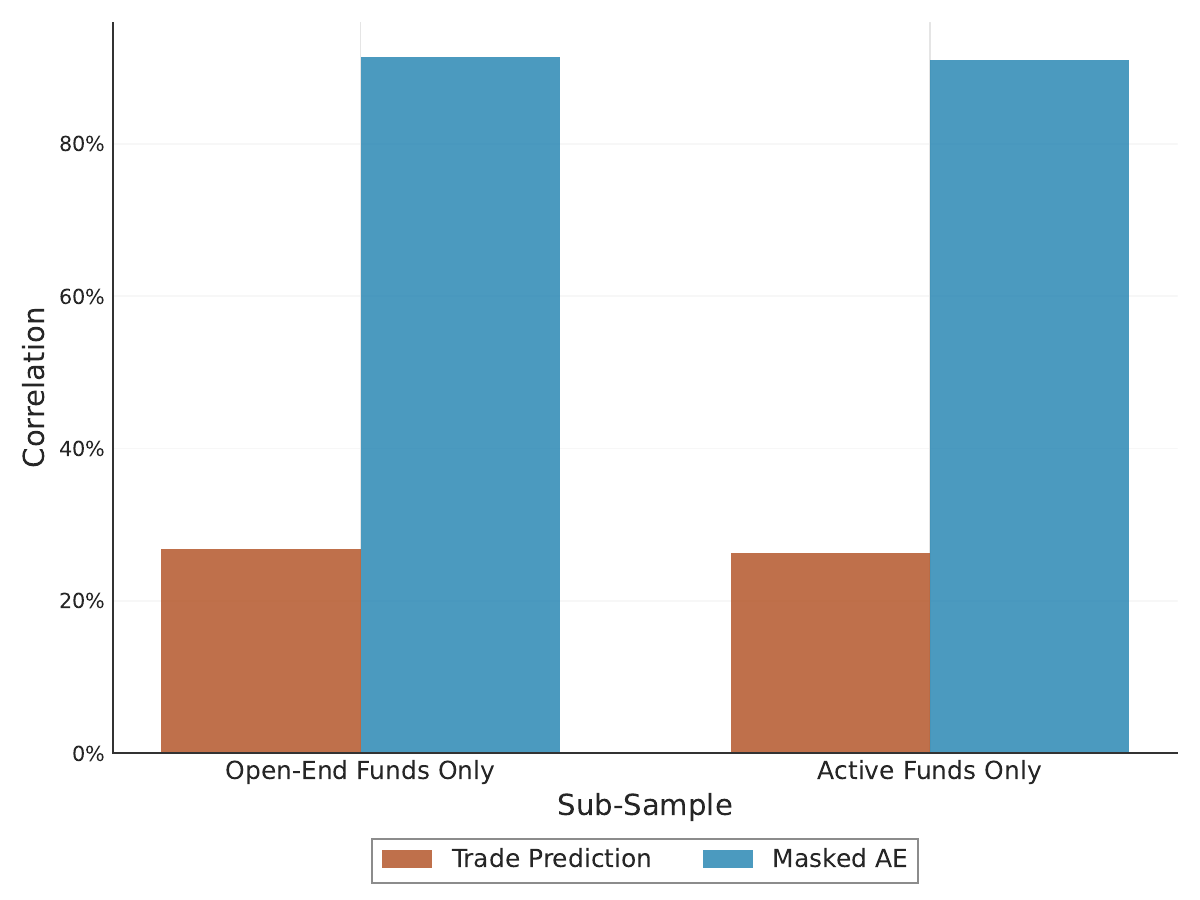}
        \label{fig:sub1}
    \end{subfigure}
    \hfill
    \begin{subfigure}{0.65\textwidth}
        \centering
        \caption{Performance for trade prediction, by date}\label{fig:het-b}
        \includegraphics[width=\textwidth]{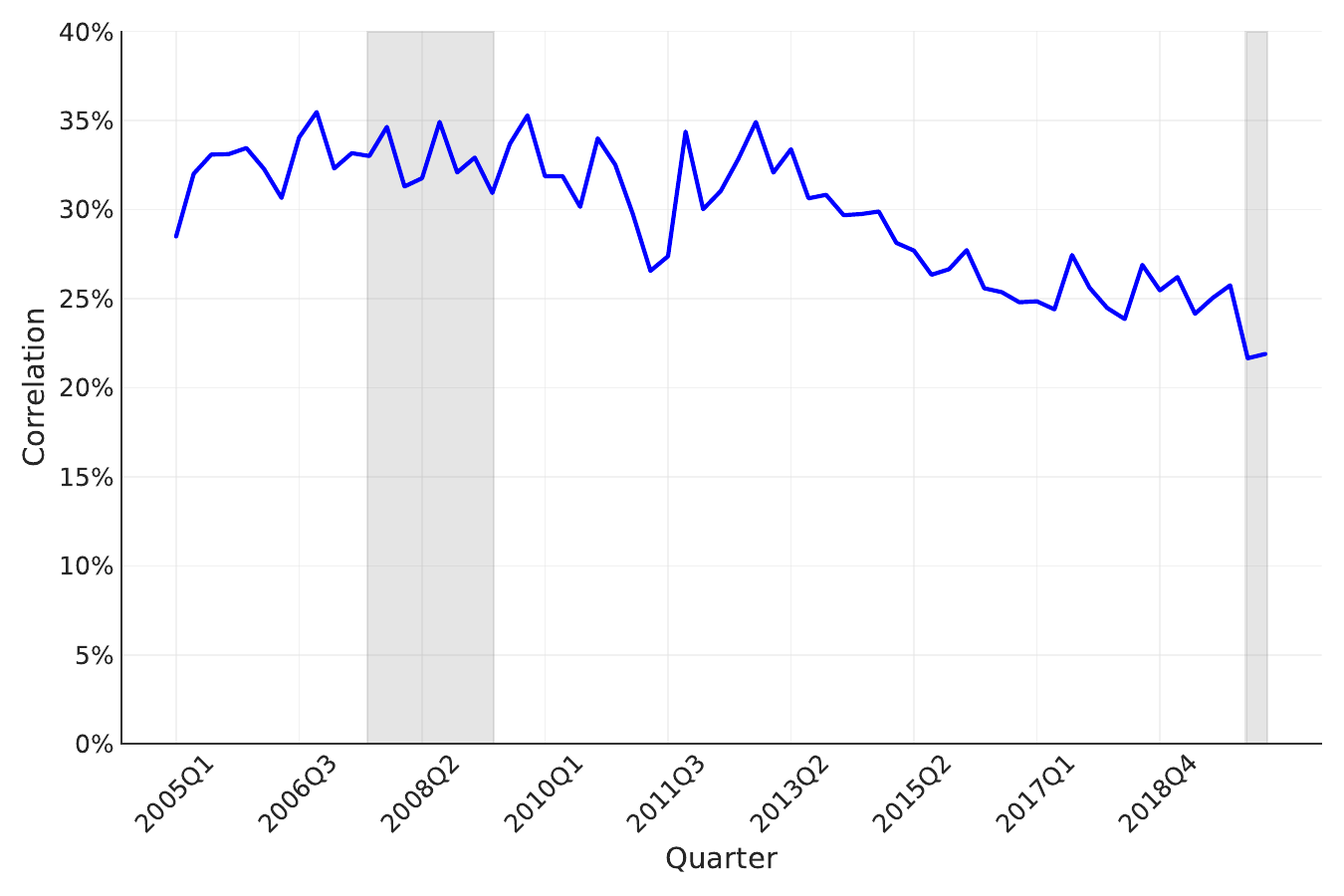}
        \label{fig:sub2}
    \end{subfigure}
    \end{center}
    \small \textit{Notes:} Panel A shows the out-of-sample correlation on the validation set between the trained model's predictions and the targets, for both tasks (trade prediction and masked autoencoder). We show this in the subsamples consisting of open-end mutual funds only and of active funds only. Panel B shows the correlation between trade predictions and targets on the full training and validation data, by quarter. Shaded gray areas correspond to periods where the St.\ Louis Fed Financial Stress Index is above 1.5.
\end{figure}

A natural question is whether the predictive ability of the model comes primarily from relatively more mechanical aspects of the data, such as by correctly assessing the trades of passive index-tracking investors. In Figure \ref{fig:het-a}, we show that this is not the case by reporting the out-of-sample performance of the model on a sub-sample consisting only of active investors. We also show performance on a sub-sample that only includes open-end mutual funds, as these are the investor category with the highest degree of coverage within the Factset holdings data. In both cases, the forecasting performance of the model is quantitatively similar to that on the full validation sample. 

An additional possible concern is that the model's predictive ability may be concentrated in calm periods rather than the market stress episodes where macroprudential interventions are most relevant. To rule this out, in Figure \ref{fig:het-b} we also show the performance on the trade prediction task separately for each quarter in the sample, as a time series: since the train-validation split occurs at the level of quarters, this time series naturally combines both training and validation data. The shaded gray areas correspond to market stress periods, defined as those when the St.\ Louis Fed Financial Stress Index is above 1.5. The predictive performance is consistently high both in stress periods and in non-stress periods.\footnote{The FRED ticker for the St.\ Louis Fed Financial Stress Index is STLFSI4. We also note that the model's performance exhibits a slight downtrend occurring between 2012 and 2017, stabilizing by the end of the period.} In particular, the model's performance during the Covid crisis yields a particularly stringent test, since none of the quarters following 2019Q3 are included in the training set: the model is only trained with pre-crisis data, precisely as it would be deployed in an actual regulatory scenario, and nonetheless it displays high accuracy in predicting the patterns of trading during the course of the crisis quarters. Altogether, these results demonstrate the efficacy of graph‐based predictive models for real-time macroprudential surveillance.

\section{Conclusion}

This paper develops a theoretical and empirical framework for understanding the role of high-dimensional predictive models in financial regulation. We formalize the tradeoffs regulators face when deploying models that deliver precise forecasts but limited causal insight, and we characterize when and how such models can improve welfare. We introduce a graph-based deep learning architecture tailored to holdings data, which we use to learn representations of assets and investors which achieve state-of-the-art results in forecasting trading patterns with minimal out-of-sample performance loss. Our empirical analysis demonstrates that real-time prediction of portfolio dynamics is feasible and provides a blueprint for practical implementation. While predictive models are not substitutes for structural content or causal inference, our results suggest that they can meaningfully complement structural knowledge, particularly when regulators have the ability to target ex post interventions.

\newpage

\fontsize{11}{11}\selectfont
\setstretch{1.5}
\setlength{\bibsep}{5pt}
\bibliographystyle{aer-nodash}
\bibliography{fsai.bib}
\newpage{}
\fontsize{11}{13}\selectfont

\end{document}

%% file: graphs/architecture_tikz.tex
\begin{tikzpicture}[
    node distance=1.5cm,
    layer/.style={draw=gray!30, fill=gray!5, rounded corners=8pt, minimum width=14cm, minimum height=2.5cm},
    stackedlayer/.style={draw=gray!30, fill=gray!5, rounded corners=8pt, minimum width=14cm, minimum height=3cm},
    transform/.style={draw=blue!60, fill=blue!10, rounded corners=4pt, minimum width=2.5cm, minimum height=0.8cm, font=\small},
    embedding/.style={draw=orange!70, fill=orange!15, rounded corners=4pt, minimum width=2.5cm, minimum height=0.8cm, font=\small},
    head/.style={draw=green!60, fill=green!10, rounded corners=4pt, minimum width=3cm, minimum height=1cm, font=\small\bfseries},
    investor/.style={circle, draw=blue!70, fill=blue!20, minimum size=0.5cm},
    asset/.style={circle, draw=purple!70, fill=purple!20, minimum size=0.5cm},
    attention/.style={->, >=stealth, dashed, blue!50, thick},
    flow/.style={->, >=stealth, thick, gray!70},
    bipartite/.style={gray!30, thin},
    loop/.style={->, >=stealth, thick, blue!60}
]
\node[layer, minimum height=1.8cm] (input-layer) at (0,-5) {};
\node[font=\footnotesize\itshape, gray, anchor=west] at (-6.8,-5.7) {Input Layer};
\node[layer, minimum height=2cm] (embed-layer) at (0,-2.5) {};
\node[font=\footnotesize\itshape, gray, anchor=west] at (-6.8,-3.3) {Feature Embedding};
\node[stackedlayer, minimum height=4cm] (mp-layer) at (0,1.1) {};
\node[font=\footnotesize\itshape, gray, anchor=west] at (-6.8,-0.4) {Message Passing Layers};
\node[layer, minimum height=2cm] (final-layer) at (0,4.5) {};
\node[font=\footnotesize\itshape, gray, anchor=west] at (-6.8,3.7) {Final Embeddings};
\foreach \i in {1,2,3} {
    \node[investor] (i\i) at (-3,-5) {};
}
\node[below=1mm of i2, font=\footnotesize] {Investor Nodes};
\foreach \j in {1,2,3} {
    \node[asset] (a\j) at (3,-5) {};
}
\node[below=1mm of a2, font=\footnotesize] {Asset Nodes};
\draw[bipartite] (i1) -- (a1);
\draw[bipartite] (i1) -- (a2);
\draw[bipartite] (i2) -- (a2);
\draw[bipartite] (i2) -- (a3);
\draw[bipartite] (i3) -- (a1);
\draw[bipartite] (i3) -- (a3);
\node[transform] (phi-i) at (-3,-2.5) {$\phi(x_i)$};
\node[transform] (phi-a) at (3,-2.5) {$\phi(x_a)$};
\draw[flow] (-3,-4.5) -- (phi-i);
\draw[flow] (3,-4.5) -- (phi-a);
\node[transform] (mp-i) at (-3,0.4) {$h^{(\ell)}_i$};
\node[transform] (mp-a) at (3,0.4) {$h^{(\ell)}_a$};

\node[transform, draw=purple!60, fill=purple!10] (attention) at (0,0.4) {Attention};
\node[transform, draw=green!60, fill=green!10] (message) at (0,1.6) {$m^{(\ell)}_v$};
\node[font=\scriptsize, gray, above=1mm of message] {Messages};

\node[transform, draw=blue!40, fill=blue!5] (mp-i-new) at (-3,2.6) {$h^{(\ell+1)}_i$};
\node[transform, draw=blue!40, fill=blue!5] (mp-a-new) at (3,2.6) {$h^{(\ell+1)}_a$};

\draw[dashed, gray!20, thin] (mp-i) -- (mp-i-new);
\draw[dashed, gray!20, thin] (mp-a) -- (mp-a-new);

\draw[attention] (mp-i) -- (attention);
\draw[attention] (mp-a) -- (attention);

\draw[flow] (attention) -- (message);

\draw[flow, gray] (mp-i.north) to[out=90,in=220] (message.south west);
\draw[flow, gray] (mp-a.north) to[out=90,in=320] (message.south east);

\draw[flow, gray] (message) -| (mp-i-new);
\draw[flow, gray] (message) -| (mp-a-new);
\draw[loop] ($(mp-layer.east)+(0,1.3)$) -- ++(0.8,0) -- ++(0,-2.6) -- ++(-0.8,0);

\path ($(mp-layer.west)+(0,1.3)$) -- ++(-0.8,0) -- ++(0,-2.6) -- ++(0.8,0);
\draw[flow] (phi-i) -- (mp-i);
\draw[flow] (phi-a) -- (mp-a);
\node[embedding] (e-i) at (-3,4.5) {$e_i = \rho(h^{(L)}_i)$};
\node[embedding] (e-a) at (3,4.5) {$e_a = \rho(h^{(L)}_a)$};
\draw[flow] (mp-i-new) -- (e-i);
\draw[flow] (mp-a-new) -- (e-a);
\node[head] (mae-head) at (-3.2,6.5) {\footnotesize Masked Autoencoder $\hat{w}_{ia} = f_{AE}(e_i, e_a)$};
\node[head] (tp-head) at (3.5,6.5) {\footnotesize Trade Prediction $\hat{y}_{ia} = f_{TP}(e_i, e_a)$};

\coordinate (junction-left) at (-0.5, 5.8);
\coordinate (junction-right) at (0.5, 5.8);

\draw[flow] (e-i) |- (junction-left);
\draw[flow] (e-a) |- (junction-right);

\draw[flow] (junction-left) -| (mae-head.south);
\draw[flow] (junction-right) -| (mae-head.south);
\draw[flow] (junction-left) -| (tp-head.south);
\draw[flow] (junction-right) -| (tp-head.south);
\end{tikzpicture}

%% file: fsai.bib
@article{maron2018invariant,
  title={Invariant and equivariant graph networks},
  author={Maron, Haggai and Ben-Hamu, Heli and Shamir, Nadav and Lipman, Yaron},
  journal={arXiv preprint arXiv:1812.09902},
  year={2018}
}

@article{zaheer2017deep,
  title={Deep sets},
  author={Zaheer, Manzil and Kottur, Satwik and Ravanbakhsh, Siamak and Poczos, Barnabas and Salakhutdinov, Russ R and Smola, Alexander J},
  journal={Advances in neural information processing systems},
  volume={30},
  year={2017}
}

@article{laffont1986,
 author = {Jean-Jacques Laffont and Jean Tirole},
 journal = {Journal of Political Economy},
 number = {3},
 pages = {614--641},
 publisher = {University of Chicago Press},
 title = {Using Cost Observation to Regulate Firms},
 urldate = {2025-07-22},
 volume = {94},
 year = {1986}
}

@misc{bernanke1986agency,
  title={Agency costs, collateral, and business fluctuations},
  author={Bernanke, Ben S and Gertler, Mark},
  year={1986},
  publisher={National Bureau of Economic Research Cambridge, Mass., USA}
}

@article{williamson1988corporate,
  title={Corporate finance and corporate governance},
  author={Williamson, Oliver E},
  journal={The journal of finance},
  volume={43},
  number={3},
  pages={567--591},
  year={1988},
  publisher={Wiley Online Library}
}

@article{fontanier2025,
title = {Optimal policy for behavioral financial crises},
journal = {Journal of Financial Economics},
volume = {166},
pages = {104005},
year = {2025},
issn = {0304-405X},
doi = {https://doi.org/10.1016/j.jfineco.2025.104005},
url = {https://www.sciencedirect.com/science/article/pii/S0304405X25000133},
author = {Paul Fontanier}
}

@article{elliott2014financial,
  title={Financial networks and contagion},
  author={Elliott, Matthew and Golub, Benjamin and Jackson, Matthew O},
  journal={American Economic Review},
  volume={104},
  number={10},
  pages={3115--3153},
  year={2014},
  publisher={American Economic Association 2014 Broadway, Suite 305, Nashville, TN 37203}
}

@article{acemoglu2015systemic,
  title={Systemic risk and stability in financial networks},
  author={Acemoglu, Daron and Ozdaglar, Asuman and Tahbaz-Salehi, Alireza},
  journal={American Economic Review},
  volume={105},
  number={2},
  pages={564--608},
  year={2015},
  publisher={American Economic Association 2014 Broadway, Suite 305, Nashville, TN 37203}
}

@article{athey2021policy,
  title={Policy learning with observational data},
  author={Athey, Susan and Wager, Stefan},
  journal={Econometrica},
  volume={89},
  number={1},
  pages={133--161},
  year={2021},
  publisher={Wiley Online Library}
}

@article{barro1983positive,
  title={A positive theory of monetary policy in a natural rate model},
  author={Barro, Robert J and Gordon, David B},
  journal={Journal of political economy},
  volume={91},
  number={4},
  pages={589--610},
  year={1983},
  publisher={The University of Chicago Press}
}

@article{brainard1968pitfalls,
  title={Pitfalls in financial model building},
  author={Brainard, William C and Tobin, James},
  journal={The American economic review},
  volume={58},
  number={2},
  pages={99--122},
  year={1968},
  publisher={JSTOR}
}

@book{kindleberger1978manias,
  title={Manias, panics and crashes: a history of financial crises},
  author={Kindleberger, Charles and Aliber, Robert},
  year={1978},
  publisher={Springer}
}

@article{fisher1933debt,
  title={The debt-deflation theory of great depressions},
  author={Fisher, Irving},
  journal={Econometrica: Journal of the Econometric Society},
  pages={337--357},
  year={1933},
  publisher={JSTOR}
}

@inproceedings{lucas1976econometric,
  title={Econometric policy evaluation: A critique},
  author={Lucas, Robert E},
  booktitle={Carnegie-Rochester conference series on public policy},
  volume={1},
  pages={19--46},
  year={1976},
  organization={North-Holland}
}

@article{friedman1953methodology,
  title={The methodology of positive economics},
  author={Friedman, Milton},
  year={1953},
  journal={Essays In Positive Economics},
  publisher={Chicago}
}

@article{haavelmo1944probability,
  title={The probability approach in econometrics},
  author={Haavelmo, Trygve},
  journal={Econometrica: Journal of the Econometric Society},
  pages={iii--115},
  year={1944},
  publisher={JSTOR}
}

@book{burns1946measuring,
  title={Measuring business cycles},
  author={Burns, Arthur F and Mitchell, Wesley C},
  year={1946},
  publisher={National bureau of economic research}
}

@article{koopmans1947measurement,
  title={Measurement without theory},
  author={Koopmans, Tjalling C},
  journal={The Review of Economics and Statistics},
  volume={29},
  number={3},
  pages={161--172},
  year={1947},
  publisher={JSTOR}
}

@article{kingma2014adam,
  title={Adam: A method for stochastic optimization},
  author={Kingma, Diederik P},
  journal={arXiv preprint arXiv:1412.6980},
  year={2014}
}

@article{hendrycks2016gaussian,
  title={Gaussian error linear units (gelus)},
  author={Hendrycks, Dan and Gimpel, Kevin},
  journal={arXiv preprint arXiv:1606.08415},
  year={2016}
}

@article{jiang2021could,
  title={Could graph neural networks learn better molecular representation for drug discovery? A comparison study of descriptor-based and graph-based models},
  author={Jiang, Dejun and Wu, Zhenxing and Hsieh, Chang-Yu and Chen, Guangyong and Liao, Ben and Wang, Zhe and Shen, Chao and Cao, Dongsheng and Wu, Jian and Hou, Tingjun},
  journal={Journal of cheminformatics},
  volume={13},
  pages={1--23},
  year={2021},
  publisher={Springer}
}

@inproceedings{derrow2021eta,
  title={Eta prediction with graph neural networks in google maps},
  author={Derrow-Pinion, Austin and She, Jennifer and Wong, David and Lange, Oliver and Hester, Todd and Perez, Luis and Nunkesser, Marc and Lee, Seongjae and Guo, Xueying and Wiltshire, Brett and others},
  booktitle={Proceedings of the 30th ACM international conference on information \& knowledge management},
  pages={3767--3776},
  year={2021}
}

@article{kozak2020shrinking,
  title={Shrinking the cross-section},
  author={Kozak, Serhiy and Nagel, Stefan and Santosh, Shrihari},
  journal={Journal of Financial Economics},
  volume={135},
  number={2},
  pages={271--292},
  year={2020},
  publisher={Elsevier}
}

@article{vaswani2017attention,
  title={Attention is all you need},
  author={Vaswani, Ashish and Shazeer, Noam and Parmar, Niki and Uszkoreit, Jakob and Jones, Llion and Gomez, Aidan N and Kaiser, {\L}ukasz and Polosukhin, Illia},
  journal={Advances in neural information processing systems},
  volume={30},
  year={2017}
}

@article{brunnermeier2013bubbles,
  title={Bubbles, financial crises, and systemic risk},
  author={Brunnermeier, Markus K and Oehmke, Martin},
  journal={Handbook of the Economics of Finance},
  volume={2},
  pages={1221--1288},
  year={2013},
  publisher={Elsevier}
}

@article{krishnamurthy2025credit,
  title={How credit cycles across a financial crisis},
  author={Krishnamurthy, Arvind and Muir, Tyler},
  journal={The Journal of Finance},
  volume={80},
  number={3},
  pages={1339--1378},
  year={2025},
  publisher={Wiley Online Library}
}

@article{haddad2021selling,
  title={When selling becomes viral: Disruptions in debt markets in the COVID-19 crisis and the Fed’s response},
  author={Haddad, Valentin and Moreira, Alan and Muir, Tyler},
  journal={The Review of Financial Studies},
  volume={34},
  number={11},
  pages={5309--5351},
  year={2021},
  publisher={Oxford University Press}
}

@article{adrian2014procyclical,
  title={Procyclical leverage and value-at-risk},
  author={Adrian, Tobias and Shin, Hyun Song},
  journal={The Review of Financial Studies},
  volume={27},
  number={2},
  pages={373--403},
  year={2014},
  publisher={Oxford University Press}
}

@article{coval2007asset,
  title={Asset fire sales (and purchases) in equity markets},
  author={Coval, Joshua and Stafford, Erik},
  journal={Journal of Financial Economics},
  volume={86},
  number={2},
  pages={479--512},
  year={2007},
  publisher={Elsevier}
}

@article{fang2025holds,
  title={Who holds sovereign debt and why it matters},
  author={Fang, Xiang and Hardy, Bryan and Lewis, Karen K},
  journal={The Review of Financial Studies},
  pages={hhaf031},
  year={2025},
  publisher={Oxford University Press}
}

@article{coppola2025safe,
  title={In safe hands: The financial and real impact of investor composition over the credit cycle},
  author={Coppola, Antonio},
  journal={The Review of Financial Studies},
  pages={hhaf017},
  year={2025},
  publisher={Oxford University Press}
}

@article{schularick2012credit,
  title={Credit booms gone bust: monetary policy, leverage cycles, and financial crises, 1870--2008},
  author={Schularick, Moritz and Taylor, Alan M},
  journal={American Economic Review},
  volume={102},
  number={2},
  pages={1029--1061},
  year={2012},
  publisher={American Economic Association}
}

@article{scarselli2008graph,
  title={The graph neural network model},
  author={Scarselli, Franco and Gori, Marco and Tsoi, Ah Chung and Hagenbuchner, Markus and Monfardini, Gabriele},
  journal={IEEE transactions on neural networks},
  volume={20},
  number={1},
  pages={61--80},
  year={2008},
  publisher={IEEE}
}

@article{xu2018powerful,
  title={How powerful are graph neural networks?},
  author={Xu, Keyulu and Hu, Weihua and Leskovec, Jure and Jegelka, Stefanie},
  journal={arXiv preprint arXiv:1810.00826},
  year={2018}
}

@article{hamilton2017inductive,
  title={Inductive representation learning on large graphs},
  author={Hamilton, Will and Ying, Zhitao and Leskovec, Jure},
  journal={Advances in neural information processing systems},
  volume={30},
  year={2017}
}

@article{farboodi2020long,
  title={Long-run growth of financial data technology},
  author={Farboodi, Maryam and Veldkamp, Laura},
  journal={American Economic Review},
  volume={110},
  number={8},
  pages={2485--2523},
  year={2020},
  publisher={American Economic Association 2014 Broadway, Suite 305, Nashville, TN 37203}
}

@article{bryzgalova2023asset,
  title={Asset-pricing factors with economic targets},
  author={Bryzgalova, Svetlana and DeMiguel, Victor and Li, Sicong and Pelger, Markus},
  journal={Available at SSRN},
  volume={4344837},
  year={2023}
}

@article{leitner2023model,
  title={Model secrecy and stress tests},
  author={Leitner, Yaron and Williams, Basil},
  journal={The Journal of Finance},
  volume={78},
  number={2},
  pages={1055--1095},
  year={2023},
  publisher={Wiley Online Library}
}

@article{orlov2023design,
  title={The design of macroprudential stress tests},
  author={Orlov, Dmitry and Zryumov, Pavel and Skrzypacz, Andrzej},
  journal={The Review of Financial Studies},
  volume={36},
  number={11},
  pages={4460--4501},
  year={2023},
  publisher={Oxford University Press}
}

@article{faria2017runs,
  title={Runs versus lemons: Information disclosure and fiscal capacity},
  author={Faria-e-Castro, Miguel and Martinez, Joseba and Philippon, Thomas},
  journal={The Review of Economic Studies},
  volume={84},
  number={4},
  pages={1683--1707},
  year={2017},
  publisher={Oxford University Press}
}

@article{shapiro2015information,
  title={Information management in banking crises},
  author={Shapiro, Joel and Skeie, David},
  journal={The Review of Financial Studies},
  volume={28},
  number={8},
  pages={2322--2363},
  year={2015},
  publisher={Oxford University Press}
}

@article{parlatore2025designing,
  title={Designing stress scenarios},
  author={Parlatore, Cecilia and Philippon, Thomas},
  journal={The Journal of Finance},
  volume={80},
  number={2},
  pages={833--873},
  year={2025},
  publisher={Wiley Online Library}
}

@article{goldstein2018stress,
  title={Stress tests and information disclosure},
  author={Goldstein, Itay and Leitner, Yaron},
  journal={Journal of Economic Theory},
  volume={177},
  pages={34--69},
  year={2018},
  publisher={Elsevier}
}

@article{kleinberg2018human,
  title={Human decisions and machine predictions},
  author={Kleinberg, Jon and Lakkaraju, Himabindu and Leskovec, Jure and Ludwig, Jens and Mullainathan, Sendhil},
  journal={The quarterly journal of economics},
  volume={133},
  number={1},
  pages={237--293},
  year={2018},
  publisher={Oxford University Press}
}

@incollection{athey2018impact,
  title={The impact of machine learning on economics},
  author={Athey, Susan},
  booktitle={The economics of artificial intelligence: An agenda},
  pages={507--547},
  year={2018},
  publisher={University of Chicago Press}
}

@article{athey2017beyond,
  title={Beyond prediction: Using big data for policy problems},
  author={Athey, Susan},
  journal={Science},
  volume={355},
  number={6324},
  pages={483--485},
  year={2017},
  publisher={American Association for the Advancement of Science}
}

@article{kleinberg2015prediction,
  title={Prediction policy problems},
  author={Kleinberg, Jon and Ludwig, Jens and Mullainathan, Sendhil and Obermeyer, Ziad},
  journal={American Economic Review},
  volume={105},
  number={5},
  pages={491--495},
  year={2015},
  publisher={American Economic Association 2014 Broadway, Suite 305, Nashville, TN 37203}
}

@article{gillis2019big,
  title={Big data and discrimination},
  author={Gillis, Talia B and Spiess, Jann L},
  journal={The University of Chicago Law Review},
  volume={86},
  number={2},
  pages={459--488},
  year={2019},
  publisher={JSTOR}
}

@article{mullainathan2017machine,
  title={Machine learning: an applied econometric approach},
  author={Mullainathan, Sendhil and Spiess, Jann},
  journal={Journal of Economic Perspectives},
  volume={31},
  number={2},
  pages={87--106},
  year={2017},
  publisher={American Economic Association 2014 Broadway, Suite 305, Nashville, TN 37203-2418}
}

@article{chen2024deep,
  title={Deep learning in asset pricing},
  author={Chen, Luyang and Pelger, Markus and Zhu, Jason},
  journal={Management Science},
  volume={70},
  number={2},
  pages={714--750},
  year={2024},
  publisher={INFORMS}
}

@article{giglio2022factor,
  title={Factor models, machine learning, and asset pricing},
  author={Giglio, Stefano and Kelly, Bryan and Xiu, Dacheng},
  journal={Annual Review of Financial Economics},
  volume={14},
  number={1},
  pages={337--368},
  year={2022},
  publisher={Annual Reviews}
}

@article{einav2014data,
  title={The data revolution and economic analysis},
  author={Einav, Liran and Levin, Jonathan},
  journal={Innovation Policy and the Economy},
  volume={14},
  number={1},
  pages={1--24},
  year={2014},
  publisher={University of Chicago Press Chicago, IL}
}

@article{einav2014economics,
  title={Economics in the age of big data},
  author={Einav, Liran and Levin, Jonathan},
  journal={Science},
  volume={346},
  number={6210},
  pages={1243089},
  year={2014},
  publisher={American Association for the Advancement of Science}
}

@misc{sarkar2025economic,
  title={Economic representations},
  author={Sarkar, Suproteem K},
  year={2025},
  publisher={jan}
}

@article{gabaix2025upgrading,
  title={Upgrading Credit Pricing and Risk Assessment through Embeddings},
  author={Gabaix, Xavier and Koijen, Ralph SJ and Richmond, Robert and Yogo, Motohiro},
  journal={Available at SSRN},
  year={2025}
}

@article{dolphin2022stock,
  title={Stock embeddings: Learning distributed representations for financial assets},
  author={Dolphin, Rian and Smyth, Barry and Dong, Ruihai},
  journal={arXiv preprint arXiv:2202.08968},
  year={2022}
}

@article{gabaix2024asset,
  title={Asset embeddings},
  author={Gabaix, Xavier and Koijen, Ralph SJ and Richmond, Robert and Yogo, Motohiro},
  journal={Available at SSRN 4507511},
  year={2024}
}

@book{nagel2021machine,
  title={Machine learning in asset pricing},
  author={Nagel, Stefan},
  year={2021},
  publisher={Princeton University Press}
}

@article{gu2021autoencoder,
  title={Autoencoder asset pricing models},
  author={Gu, Shihao and Kelly, Bryan and Xiu, Dacheng},
  journal={Journal of Econometrics},
  volume={222},
  number={1},
  pages={429--450},
  year={2021},
  publisher={Elsevier}
}

@article{gu2020empirical,
  title={Empirical asset pricing via machine learning},
  author={Gu, Shihao and Kelly, Bryan and Xiu, Dacheng},
  journal={The Review of Financial Studies},
  volume={33},
  number={5},
  pages={2223--2273},
  year={2020},
  publisher={Oxford University Press}
}

@article{wu2020comprehensive,
  title={A comprehensive survey on graph neural networks},
  author={Wu, Zonghan and Pan, Shirui and Chen, Fengwen and Long, Guodong and Zhang, Chengqi and Yu, Philip S},
  journal={IEEE transactions on neural networks and learning systems},
  volume={32},
  number={1},
  pages={4--24},
  year={2020},
  publisher={IEEE}
}

@article{jumper2021highly,
  title={Highly accurate protein structure prediction with AlphaFold},
  author={Jumper, John and Evans, Richard and Pritzel, Alexander and Green, Tim and Figurnov, Michael and Ronneberger, Olaf and Tunyasuvunakool, Kathryn and Bates, Russ and {\v{Z}}{\'\i}dek, Augustin and Potapenko, Anna and others},
  journal={nature},
  volume={596},
  number={7873},
  pages={583--589},
  year={2021},
  publisher={Nature Publishing Group}
}

@article{bianchi2016,
Author = {Bianchi, Javier},
Title = {Efficient Bailouts?},
Journal = {American Economic Review},
Volume = {106},
Number = {12},
Year = {2016},
Month = {December},
Pages = {3607–59},
DOI = {10.1257/aer.20121524},
URL = {https://www.aeaweb.org/articles?id=10.1257/aer.20121524}}

@article{claytonschaab2025,
    author = {Clayton, Christopher and Schaab, Andreas},
    title = {Bail-Ins, Optimal Regulation, and Crisis Resolution},
    journal = {The Review of Financial Studies},
    pages = {hhaf002},
    year = {2025},
    month = {03},
    issn = {0893-9454},
    doi = {10.1093/rfs/hhaf002},
    url = {https://doi.org/10.1093/rfs/hhaf002},
    eprint = {https://academic.oup.com/rfs/advance-article-pdf/doi/10.1093/rfs/hhaf002/62444185/hhaf002.pdf},
}

@article{davilakorinek2018,
    author = {Dávila, Eduardo and Korinek, Anton},
    title = {Pecuniary Externalities in Economies with Financial Frictions},
    journal = {The Review of Economic Studies},
    volume = {85},
    number = {1},
    pages = {352-395},
    year = {2017},
    month = {02},
    issn = {0034-6527},
    doi = {10.1093/restud/rdx010},
    url = {https://doi.org/10.1093/restud/rdx010},
    eprint = {https://academic.oup.com/restud/article-pdf/85/1/352/23033540/rdx010.pdf},
}

@article{bianchimendoza2016,
author = {Bianchi, Javier and Mendoza, Enrique G.},
title = {Optimal Time-Consistent Macroprudential Policy},
journal = {Journal of Political Economy},
volume = {126},
number = {2},
pages = {588-634},
year = {2018},
doi = {10.1086/696280},

URL = { 
    
        https://doi.org/10.1086/696280
    
    

},
eprint = { 
    
        https://doi.org/10.1086/696280
    
    

}
}

@article{charikehoe2016,
Author = {Chari, V. V. and Kehoe, Patrick J.},
Title = {Bailouts, Time Inconsistency, and Optimal Regulation: A Macroeconomic View},
Journal = {American Economic Review},
Volume = {106},
Number = {9},
Year = {2016},
Month = {September},
Pages = {2458–93},
DOI = {10.1257/aer.20150157},
URL = {https://www.aeaweb.org/articles?id=10.1257/aer.20150157}}

@article{stein2012,
    author = {Stein, Jeremy C.},
    title = {Monetary Policy as Financial Stability Regulation*},
    journal = {The Quarterly Journal of Economics},
    volume = {127},
    number = {1},
    pages = {57-95},
    year = {2012},
    month = {01},
    issn = {0033-5533},
    doi = {10.1093/qje/qjr054},
    url = {https://doi.org/10.1093/qje/qjr054},
    eprint = {https://academic.oup.com/qje/article-pdf/127/1/57/5188446/qjr054.pdf},
}

@article{Bianchi2011,
Author = {Bianchi, Javier},
Title = {Overborrowing and Systemic Externalities in the Business Cycle},
Journal = {American Economic Review},
Volume = {101},
Number = {7},
Year = {2011},
Month = {December},
Pages = {3400–3426},
DOI = {10.1257/aer.101.7.3400},
URL = {https://www.aeaweb.org/articles?id=10.1257/aer.101.7.3400}}

@article{Lorenzoni2008,
    author = {Lorenzoni, Guido},
    title = {Inefficient Credit Booms},
    journal = {The Review of Economic Studies},
    volume = {75},
    number = {3},
    pages = {809-833},
    year = {2008},
    month = {07},
    issn = {0034-6527},
    doi = {10.1111/j.1467-937X.2008.00494.x},
    url = {https://doi.org/10.1111/j.1467-937X.2008.00494.x},
    eprint = {https://academic.oup.com/restud/article-pdf/75/3/809/18352771/75-3-809.pdf},
}

@Article{Caballero2001,
  author={Caballero, Ricardo J. and Krishnamurthy, Arvind},
  title={{International and domestic collateral constraints in a model of emerging market crises}},
  journal={Journal of Monetary Economics},
  year=2001,
  volume={48},
  number={3},
  pages={513-548},
  month={December},
  doi={},
  url={https://ideas.repec.org/a/eee/moneco/v48y2001i3p513-548.html}
}

@article{kiyotaki1997credit,
	author = {Kiyotaki, Nobuhiro and Moore, John},
	journal = {Journal of political economy},
	number = {2},
	pages = {211--248},
	publisher = {The University of Chicago Press},
	title = {Credit cycles},
	volume = {105},
	year = {1997}}

@article{farhi2016theory,
	author = {Farhi, Emmanuel and Werning, Iv{\'a}n},
	journal = {Econometrica},
	number = {5},
	pages = {1645--1704},
	publisher = {Wiley Online Library},
	title = {A theory of macroprudential policies in the presence of nominal rigidities},
	volume = {84},
	year = {2016}}

@article{claytonschaab2022,
	author = {Clayton, Christopher and Schaab, Andreas},
	doi = {10.1093/qje/qjac002},
	eprint = {https://academic.oup.com/qje/article-pdf/137/3/1681/44839859/qjac002.pdf},
	issn = {0033-5533},
	journal = {The Quarterly Journal of Economics},
	month = {01},
	number = {3},
	pages = {1681-1736},
	title = {{Multinational Banks and Financial Stability}},
	url = {https://doi.org/10.1093/qje/qjac002},
	volume = {137},
	year = {2022}}

@article{coppola2021redrawing,
	author = {Coppola, Antonio and Maggiori, Matteo and Neiman, Brent and Schreger, Jesse},
	journal = {The Quarterly Journal of Economics},
	number = {3},
	pages = {1499--1556},
	publisher = {Oxford University Press},
	title = {Redrawing the Map of Global Capital Flows: The Role of Cross-Border Financing and Tax Havens},
	volume = {136},
	year = {2021}}
